\pdfoutput=1

\documentclass[11pt]{article}

\usepackage{EMNLP2023}

\usepackage{times}
\usepackage{latexsym}
\usepackage{amsmath}
\usepackage[T1]{fontenc}    
\usepackage{url}            
\usepackage{booktabs}       
\usepackage{amsfonts}       
\usepackage{nicefrac}       
\usepackage{xcolor}         
\usepackage{graphicx} 
\usepackage{float}
\usepackage{subcaption}  
\usepackage{natbib}
\usepackage{enumitem} 
\usepackage{capt-of}
\usepackage{dblfloatfix}
\usepackage{rotating}
\usepackage{multirow}
\usepackage{pdflscape}
\usepackage{adjustbox}
\usepackage{caption}
\usepackage{needspace}
\usepackage{geometry}
\usepackage{longtable}
\usepackage[T1]{fontenc}

\usepackage[utf8]{inputenc}

\usepackage{microtype}

\usepackage{inconsolata}
\usepackage{pdfpages}

\newcommand{\TableFont}{\fontsize{8pt}{10pt}\selectfont} %

\title{SocioBench: Modeling Human Behavior in Sociological Surveys with Large Language Models}

\author{
  Jia Wang\textsuperscript{1}\textsuperscript{2},
  Ziyu Zhao\textsuperscript{1}\textsuperscript{3},
  Tingjuntao Ni\textsuperscript{1}\textsuperscript{4},
  Zhongyu Wei\textsuperscript{1}\textsuperscript{5}\thanks{~~Corresponding authors.}\\
\textsuperscript{1}Shanghai Innovation Institute,
\textsuperscript{2}Tongji University,
\textsuperscript{3}Zhejiang University,\\
\textsuperscript{4}Shanghai Jiao Tong University,
\textsuperscript{5}Fudan University\\
\texttt{jiawang@tongji.edu.cn, zywei@fudan.edu.cn}\\
}

\begin{document}
\maketitle
\begin{abstract}

Large language models (LLMs) show strong potential for simulating human social behaviors and interactions, yet lack large-scale, systematically constructed benchmarks for evaluating their alignment with real-world social attitudes. To bridge this gap, we introduce SocioBench—a comprehensive benchmark derived from the annually collected, standardized survey data of the \textit{International Social Survey Programme (ISSP)}. The benchmark aggregates over 480,000 real respondent records from more than 30 countries, spanning 10 sociological domains and over 40 demographic attributes. Our experiments indicate that LLMs achieve only 30–40\% accuracy when simulating individuals in complex survey scenarios, with statistically significant differences across domains and demographic subgroups. These findings highlight several limitations of current LLMs in survey scenarios, including insufficient individual-level data coverage, inadequate scenario diversity, and missing group-level modeling. We have open-sourced \textbf{SocioBench} at \url{https://github.com/JiaWANG-TJ/SocioBench}.

\end{abstract}

\section{Introduction}
As the LLMs advance in generating natural language ~\cite{10.1145/3605943,10.1145/3635059.3635104,10.1162/coli_a_00561}, simulating cognitive processes ~\cite{niu2024largelanguagemodelscognitive,10.1145/3613904.3642754,ren-etal-2025-large,articleAzaria,articleChen}, and engaging in complex dialogues ~\cite{mou2024agentsense,li2024finegrainedbehaviorsimulationroleplaying}, their potential applications in the social sciences are becoming increasingly evident ~\cite{anthis2025llm, aher2023using, chen2024persona}. Beyond analyzing large-scale textual data, LLMs can function as "computational agents" that simulate human behavior ~\cite{liu2024lmagentlargescalemultimodalagents,10.1145/3708985} and decision-making ~\cite{10970024,10.1145/3711032}, enabling social experiments and surveys ~\cite{zhang2025socioverseworldmodelsocial,leng2023llm,mou2024individualsocietysurveysocial} that are difficult to conduct in real-world settings due to ethical, logistical, or financial constraints~\cite{park2023generative}. Existing research has primarily focused on micro-level social capabilities such as persona consistency, linguistic style, and personality traits, or on group-level tasks like social reasoning, social bias identification, and multi-agent cooperation ~\cite{ji2025enhancing, strachan2024testing, li2023camel}. Although benchmarks such as OpinionQA ~\cite{santurkar2023whose} have made important strides in evaluating these aspects, few have systematically assessed LLMs' ability to reflect macro-level social attitudes and cross-cultural differences.

\begin{figure*}[!t]
    \centering
    \includegraphics[width=1\textwidth]{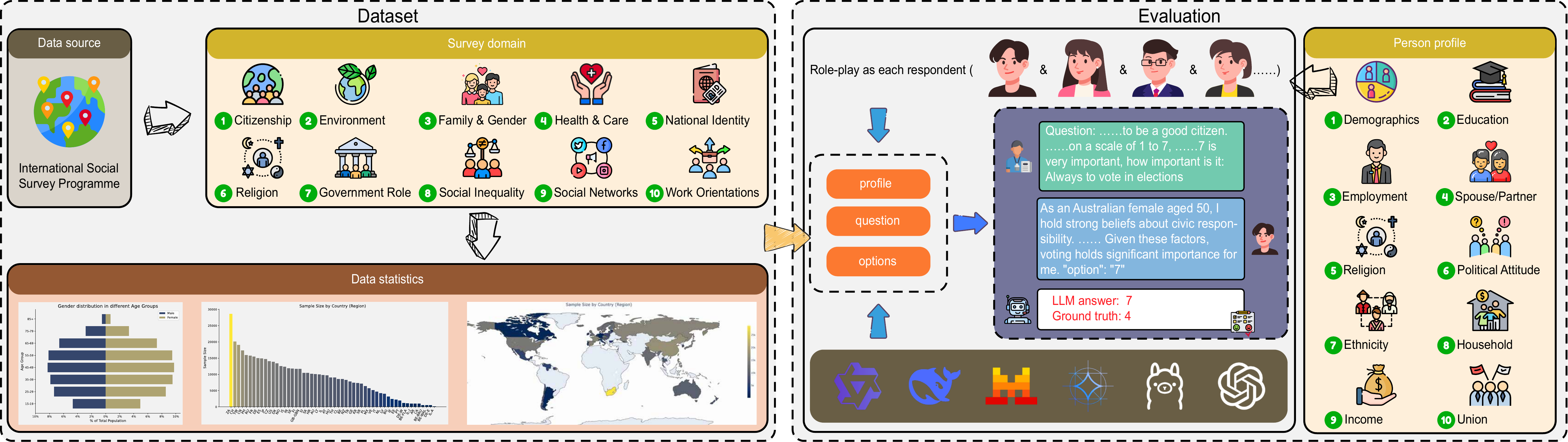}
    \caption{Overview of SocioBench. We first constructed the questionnaire question-answering dataset covering the ten sociological domains of the ISSP, along with the dataset containing ground-truth demographic labels and respondent answers. We then instructed the LLM to answer the survey conditioned on the demographic labels, and evaluated model performance by computing the accuracy between the LLM's responses and the ground-truth answers.
}
    \label{fig:SocioBench pipeline}
\end{figure*}

To bridge this gap, we develop \textbf{SocioBench}, a large-scale, cross-national benchmark for simulating human behavior in social survey scenarios. The benchmark is built upon the ISSP's ~\cite{ZA5950,ZA5900,ZA6670,ZA6770, ZA6900,ZA6980,ZA7570,ZA7600,ZA7650,ZA8000} standardized questionnaires and 481,629 authentic respondent records, and it covers 10 research domains: \textit{Citizenship, Environment, Family and Changing Gender Roles, Health and Health Care, National Identity, Religion, Role of Government, Social Inequality, Social Networks, and Work Orientations}. Figure~\ref{fig:SocioBench pipeline} shows an overview of the pipeline for constructing SocioBench.


\section{SocioBench Curation}

\paragraph{Dataset Statistics.}

SocioBench is built upon the ISSP, a long-standing, international collaborative project that annually collects standardized data on social attitudes, with its data archive maintained by the GESIS – Leibniz Institute for the Social Sciences\footnote{\url{https://www.gesis.org/en/home}}. SocioBench covers 10 sociological domains across more than 30 countries. The full version, SocioBench-Full, comprises 481,629 respondents, with each respondent profiled by over 40 demographic features—including age, gender, education level, occupation, income, religious affiliation, and political orientation et al. To enhance computational efficiency, we release sampled versions: \textbf{SocioBench-5000}, where the suffix indicates the total number of respondents. Unless otherwise specified, all experiments—excluding those in Section 4 (Data Sampling Ratios Comparison)—are conducted using SocioBench-5000. By default, "SocioBench" refers to this version. The statistical overview is presented in Table~\ref{tab:respondent-profile}, while detailed distributions of Q\&A and demographic information are available in Appendix~\ref{sec:data_stat} \& \ref{sec:Demographic_Information}.

We compare SocioBench with some representative datasets for the analysis of social attitudes and show the results in Table~\ref{tab:dataset_comparison}. Previous resources adopt partial perspectives, restricted to specific countries, a narrow set of topics, or without demographic diversity. SocioBench, on the contrary, provides a unified benchmark that simultaneously spans languages, domains, demographics, and regions, aligning more closely with real-world social contexts.

\begin{table*}[ht]
\centering
\small
\begin{tabular}{lcccc}
\toprule
\textbf{Dataset} & \textbf{Multilingual?} & \textbf{Multi-domain?} & \textbf{Demographic variables} & \textbf{Multi-regions?}  \\
\midrule
SocioBench (Ours)      & $\checkmark$ & $\checkmark$ & $\checkmark$ & $\checkmark$  \\
SocialBench ~\cite{chen2024persona} & $\times$ & $\times$ & $\times$ & $\times$  \\
OpinionQA ~\cite{santurkar2023whose}              & $\times$ & $\checkmark$ & $\checkmark$ & $\times$ \\
GlobalOpinionQA ~\cite{durmus2023towards}        & $\checkmark$ & $\checkmark$ & $\times$ & $\checkmark$ \\
\bottomrule
\end{tabular}
\caption{Comparison of social and opinion survey datasets.}
\label{tab:dataset_comparison}
\end{table*}

\begin{table}[htbp]  
  \centering
  {%
    \TableFont                    
    \setlength{\tabcolsep}{4pt}
    \begin{tabular*}{0.9\linewidth}{@{\extracolsep{\fill}}lcccccc}
      \toprule
      \textbf{Domain} & \textbf{Year} & \textbf{Ctry.} & \textbf{Feat.} &
      \textbf{Resp.} & \textbf{Q.} & \textbf{Total} \\
      \midrule
      Citizen   & 2014 & 33 & 44 & 500 & 59 & 29\,500 \\
      Enviro    & 2020 & 28 & 45 & 500 & 50 & 25\,000 \\
      Family    & 2012 & 39 & 45 & 500 & 54 & 27\,000 \\
      Health    & 2021 & 28 & 45 & 500 & 51 & 25\,500 \\
      Nat.Ident & 2013 & 35 & 46 & 500 & 60 & 30\,000 \\
      Religion  & 2018 & 30 & 46 & 500 & 59 & 29\,500 \\
      R.Gov     & 2016 & 30 & 46 & 500 & 60 & 30\,000 \\
      S.Ineq    & 2019 & 25 & 44 & 500 & 46 & 23\,000 \\
      S.Net     & 2017 & 28 & 47 & 500 & 59 & 29\,500 \\
      Work      & 2015 & 35 & 47 & 500 & 57 & 28\,500 \\
      \midrule
      \textbf{Total} & — & — & \textbf{408} & \textbf{5\,000} & \textbf{555} & \textbf{277\,500} \\
      \bottomrule
    \end{tabular*}
  } 
  \vspace{-2mm}
  \caption{Respondent profile information and questionnaire statistics in SocioBench. Abbreviations: Ctry.\ = Number of countries; Feat.\ = Number of features; Resp.\ = Number of respondents; Q.\ = Number of questions; Tot.\ = Total. Citizen = Citizenship; Enviro = Environment; Family = Family and Changing Gender Roles; Health = Health and Healthcare; Nat.Ident = National Identity; Religion = Religion; R.Gov = Role of Government; S.Ineq = Social Inequality;
    S.Net = Social Networks; Work = Work Orientations.}
  \label{tab:respondent-profile}
  \vspace{-4mm}
\end{table}

\paragraph{Dataset Curation.}

The SocioBench dataset comprises the questionnaire, respondents' demographic attributes and their responses. The data processing pipeline comprises three steps: first, we filter out open-ended questions and invalid responses (e.g., "Not applicable") in the questionnaire to retain quantifiable closed-ended items. Then, we sample 1\% of the data to form \textbf{SocioBench-5000} for experiments using a two-stage scheme—stratified by country and then random sampling within each country—in order to balance resource constraints against survey coverage. Examples from SocioBench dataset are provided in Appendix~\ref{sec:Data_example}.


\begin{table*}[ht]
  \centering
  {%
    \TableFont
    \setlength{\tabcolsep}{4pt}
    \begin{tabular*}{1\linewidth}{@{\extracolsep{\fill}}lccccccccccc}
      \toprule
      Model & Citizen & Enviro & Family & Health & Nat.Ident & Religion & R.Gov & S.Ineq & S.Net & Work & Avg. \\
      \cmidrule(lr){2-12}
      \multicolumn{1}{l}{} &
      \multicolumn{11}{c}{\textbf{\textit{Accuracy \% }($\uparrow$)}} \\
      \midrule
      \multicolumn{12}{l}{\textsc{\textbf{Baselines}}} \\
      \midrule[0.3pt]
      Random Guess & 25.93 & 23.22 & 21.58 & 21.24 & 23.02 & 20.84 & 23.64 & 20.25 & 18.65 & 22.99 & 22.14 \\
      \midrule
      GPT-4o & \textbf{44.30} & \textbf{37.07} & \textbf{39.14} & 35.33 & 36.35 & \underline{40.76} & \textbf{39.86} & \textbf{36.62} & 36.69 & \underline{38.94} & \underline{38.51} \\
      InternLM3-8b-instruct & 41.65 & 33.66 & 31.05 & 32.35 & 34.60 & 36.61 & 36.09 & 32.21 & 33.96 & 36.19 & 34.84 \\
      GLM-4-9b-chat & 41.81 & 33.95 & 31.96 & 34.13 & 36.53 & 37.32 & 36.03 & 34.35 & 31.86 & 38.10 & 35.60 \\
      Gemma-3-27b-it & 40.92 & 34.63 & 34.87 & 30.49 & 33.84 & 38.08 & 35.97 & 32.60 & 35.63 & 38.10 & 35.51 \\
      DeepSeek-R1-Distill-Llama-70B & \underline{44.19} & \underline{35.98} & 38.11 & \underline{36.14} & \underline{37.42} & 40.65 & \underline{39.32} & \underline{35.97} & \underline{37.38} & \textbf{39.99} & \textbf{38.52} \\
      \midrule[0.3pt]
      Mistral-7B-Instruct-v0.3 & 39.64 & 32.62 & 28.16 & 30.68 & 32.86 & 35.85 & 34.58 & 30.21 & 33.81 & 35.49 & 33.39 \\
      Mixtral-8x22B-Instruct-v0.1 & 43.10 & 34.20 & 34.40 & 32.38 & 33.29 & 37.86 & 35.89 & 33.70 & 37.35 & 35.11 & 35.73 \\
      \midrule[0.3pt]
      Llama-3.1-8B-Instruct & 40.43 & 32.11 & 31.89 & 32.21 & 33.37 & 36.99 & 35.27 & 31.47 & 34.99 & 33.39 & 34.21 \\
      Llama-3.3-70B-Instruct & 44.03 & 35.97 & \underline{38.62} & \textbf{36.16} & \textbf{38.19} & \textbf{41.26} & 39.19 & 35.73 & 36.14 & 38.80 & 38.41 \\
      \midrule[0.3pt]
      Qwen2.5-7B-Instruct & 40.90 & 29.84 & 30.10 & 31.82 & 33.67 & 36.54 & 34.80 & 30.37 & 33.34 & 32.18 & 33.35 \\
      Qwen2.5-32B-Instruct & 42.54 & 35.26 & 34.94 & 33.20 & 35.09 & 37.88 & 36.32 & 34.00 & 34.48 & 36.57 & 36.03 \\
      Qwen2.5-72B-Instruct & 43.59 & 35.51 & 36.27 & 35.90 & 34.13 & 39.80 & 36.56 & 35.17 & \textbf{38.06} & 37.38 & 37.24 \\
      \midrule[0.3pt]
      Qwen3-8B & 40.28 & 32.70 & 33.07 & 33.98 & 33.12 & 37.58 & 34.65 & 30.83 & 34.38 & 34.20 & 34.48 \\
      Qwen3-32B & 43.60 & 34.12 & 34.53 & 33.53 & 32.64 & 38.90 & 35.52 & 33.16 & 35.31 & 35.25 & 35.66 \\

      \bottomrule
    \end{tabular*}
  }
  \vspace{-2mm}
  \caption{Comparison of different LLMs across SocioBench. We report the best LLM performance in bold and the second best underlined.}
  \label{tab:Accuracy_comparison}
  \vspace{-6mm}
\end{table*}

\section{Experiment Setup}

\paragraph{Evaluation Pipeline.}
The evaluation pipeline engages LLMs in role-playing. A prompt template is designed to mimic authentic survey participation: LLMs are explicitly instructed to adopt the identity of the respondent through embedded demographic profiles (e.g., "You are a 31-year-old Australian woman with a high school to high school education completed, who has a partner, no religious affiliation, and is of Australian ethnicity", see ~\autoref{sec:LLMs Role-playing Prompt template}). The models then generate answer options accodrding to the sociocultural context.

\paragraph{Comparison Models.}
We compare state-of-the-art LLMs on SocioBench, including the GPT series, Llama series, Qwen series, Mistral series, and so on \cite{openai2024gpt4ocard,qwen2024qwen25,grattafiori2024llama3herdmodels,glm2024chatglmfamilylargelanguage,deepseekai2025deepseekr1incentivizingreasoningcapability,gemmateam2025gemma3technicalreport,jiang2024mixtralexperts}\footnote{\url{https://github.com/QwenLM/Qwen3}}%
\footnote{\url{https://github.com/InternLM/InternLM}}.



\paragraph{Evaluation Metrics.}
To evaluate the alignment of LLMs with real-world social attitudes in SocioBench, we employ the metrics: \textbf{Accuracy}. 
\textbf{Accuracy} measures the proportion of model predictions that exactly match the ground-truth responses:
\begin{equation}
    \text{Accuracy} = \frac{\sum_{i=1}^{n} \mathbb{I}(y_i^{\text{true}} = y_i^{\text{pred}})}{n} \times 100\%
\end{equation}
where $ y_i^{\text{true}} $ and $ y_i^{\text{pred}} $ denote the true and predicted responses for the $ i $-th sample respectively, $ n $ is the total number of valid samples, and $ \mathbb{I}(\cdot) $ is an indicator function that equals 1 when the condition is satisfied and 0 otherwise.

\paragraph{Implementation Details.}
The experiment leverages the vLLM framework to efficiently serve LLMs on 4 NVIDIA H100 GPUs supporting context lengths up to 10,240 tokens. Generation parameters are consistently maintained with a \textit{Temperature} of 0.5, \textit{Top P} of 0.95, \textit{Repetition Penalty} of 1.1.

\section{Experimental Results}

We conducted extensive experiments, systematically investigating the influence of various factors, including model parameter scale, model family, survey domain, dataset size, and survey rounds in different years. Furthermore, we examine how two factors—whether to enable reasoning and whether to output reasons—affect LLMs' behavioral simulation, and we conduct subgroup analyses based on different demographic information to further explore the bias of the LLM.

The core analyses and findings are presented in this section, while additional results are detailed in Appendix~\ref{sec:Supplementary experimental results and findings}.

\paragraph{Overall Experimental Results.}
Our experiments yielded four primary findings. First, when simulating individual behavior in complex social survey scenarios, the accuracy of LLMs is generally 30–40\% (see Table \ref{tab:Accuracy_comparison}). This shows the limitations of LLMs in modeling individual behavior.

\hspace*{0.5em}Second, model performance improves with increasing parameter scale. For instance, within the Qwen2.5 family, Qwen2.5-7B-Instruct, Qwen2.5-32B-Instruct, and Qwen2.5-72B-Instruct achieve average accuracies of 33.35\%, 36.03\%, and 37.24\%, respectively.

\hspace*{0.5em}Furthermore, across different model families, we find that GLM-4-9B-chat, Qwen2.5-32B-Instruct, and DeepSeek-R1-Distill-Llama-70B emerge as the top-performing models in the $<\!10$B, $\sim\!30$B, and $\sim\!70$B parameter ranges, achieving average accuracies of 35.60\%, 36.03\%, and 38.52\%, respectively.

\hspace*{0.5em}Finally, model performance varies significantly across different domains. For instance, accuracy peaks at 44.30\% in \textit{Citizenship} but is only 36.16\% in \textit{Health and Healthcare}. The consistent trend observed across different models is likely due to the uneven data distribution of LLM pre-training corpora. Data scarcity in certain domains results in disparities in the models' semantic comprehension capabilities when addressing sociological issues.

\paragraph{Subgroup Analyses.}
 To analyze biases that may arise when LLMs role-play respondents from different demographic backgrounds, we conducted subgroup analyses using representative models (the Qwen family, the Llama family, and the GPT family). We consider subgroups defined by geographic region (continent), sex, and age range. Moreover, we perform statistical tests to determine whether these labels significantly affect group-level accuracy in behavioral simulation. The detailed data are available in Appendix~\ref{sec:appendix_demo}.

\begin{figure*}[t]
    \centering
    \includegraphics[width=0.9\textwidth]{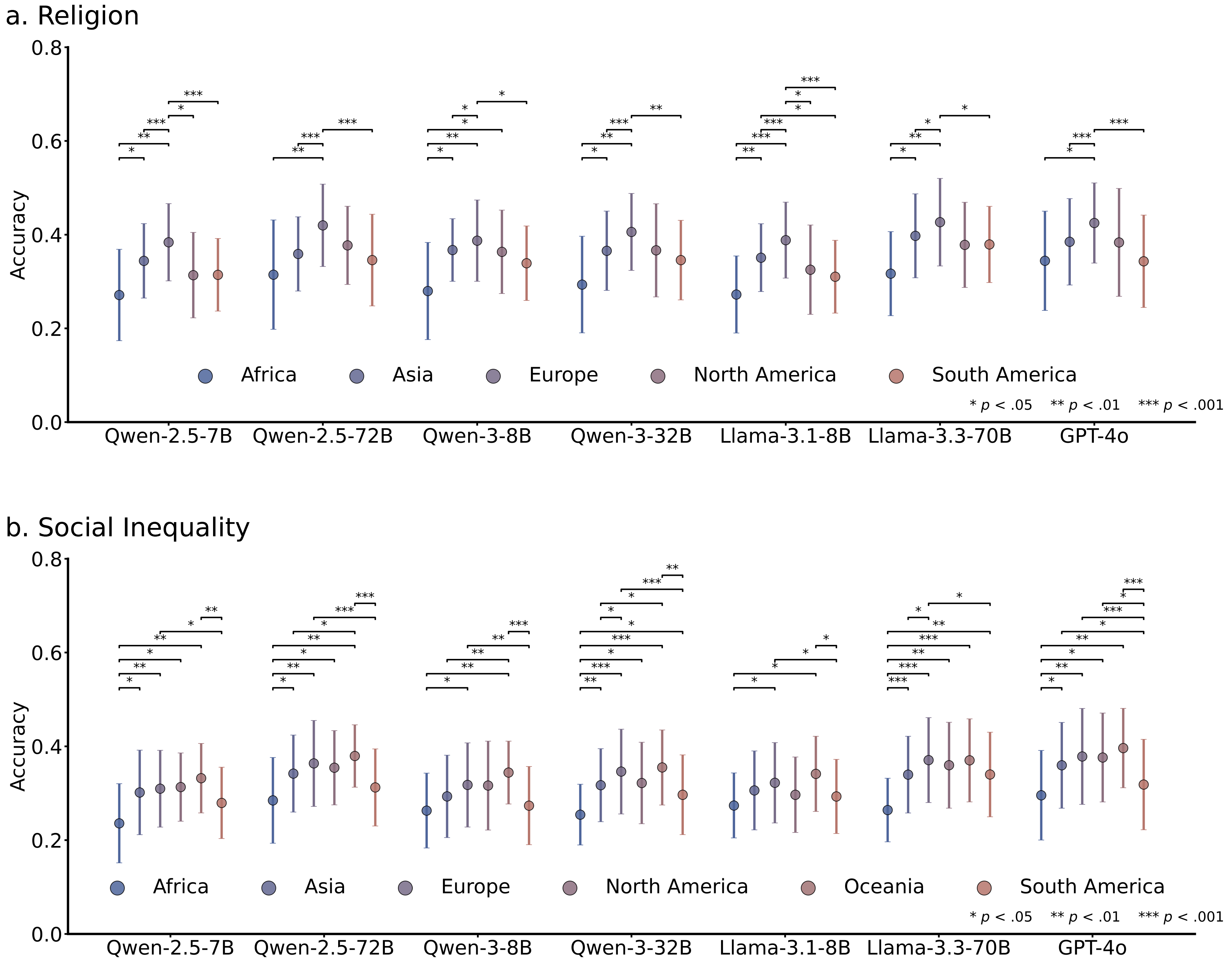}
    \caption{Experimental Results and Significance Analysis of Representative LLMs in the Cross-Continental Subgroup.}
    \label{fig:Continents_Accuracy_Comparison}
\end{figure*}

\hspace*{0.5em}\textbf{Cross-Continental Analysis: }We specifically selected the domains of \textit{Religion} and \textit{Social Inequality} for analysis, see Figure~\ref{fig:Continents_Accuracy_Comparison}. Analysis of variance reveals highly significant differences across continents for all evaluated models (all $p<.001$). Specifically, models exhibit generally lower accuracy when simulating the personas of African respondents compared to those from Europe, North America, and Oceania.

\hspace*{0.5em}\textbf{Cross-Gender Analysis: }Our analysis of the \textit{Citizenship} and \textit{Family and Changing Gender Roles} domains reveals that the accuracy in simulating female personas is consistently higher than that for male personas. For instance, the respective accuracies are 43.04\% ± 1.72\% (mean ± standard deviation) and 41.87\% ± 1.97\% in the \textit{Citizenship}. These findings suggest that training corpora imbalances may lead to female roles being associated with clearer semantic patterns in certain domains, see Figure~\ref{fig:Genders_Accuracy_Comparison}.

\hspace*{0.5em}\textbf{Cross-Age Analysis: }Our analysis shows that in the \textit{Role of Government} and \textit{Social Networks} domains, the accuracies for the 56–65 and 66-and-over age groups (37.52\% ± 2.27\% and 37.91\% ± 1.45\%, respectively) outperform young people, such as the 18–25 and 36–45 age groups. This suggests that these domains are more strongly associated with middle-aged and older populations, or that the social networks and political participation of these groups are more established, thereby enabling LLMs to simulate these demographic groups with greater accuracy, see Figure \ref{fig:Ages_Accuracy_Comparison}.

\paragraph{Option Distribution in LLMs' Responses.} 

We further conducted a comparative analysis of the distribution of options selected by human respondents and LLMs. The results reveal that although the ground truth exhibits skewed distributions (i.e., options are concentrated in several categories), the LLM-generated responses make this skewness more pronounced, and Llama-3.3-70B-Instruct shows the most marked concentration. Conversely, we observe that Qwen3-32B tends to produce more uniform option distributions. See Appendix~\ref{sec:Example-comparison} for details.

\paragraph{How do Thinking Modes Shape LLMs' Behavioral Simulation?}
To analyze how the thinking/reasoning processes affect behavioral simulation in social survey scenarios, we compared Qwen3-8B and Qwen3-32B with and without the thinking mode. The results show that the thinking mode has only a minor effect, yielding slight gains in behavioral simulation accuracy, see Table~\ref{tab:qwen3_think} in the Appendix~\ref{sec:Supplementary experimental results and findings}. Specifically, the 8B model shows an average improvement of 0.51 percentage point (pp), while the 32B model improves by 0.89 pp. An output example can be found in Appendix~\ref{sec:Comparison of Qwen3-32B With and Without Think Mode}. 


\paragraph{Data Sampling Ratios Comparison.} 

To evaluate robustness across different data scales, we further constructed two sub-datasets, SocioBench-10000 and SocioBench-20000, by sampling 2\% and 4\% of the complete dataset. On SocioBench-5000, SocioBench-10000, and SocioBench-20000, the Llama-3.1-8B-Instruct model achieved average accuracies of 34.21\%, 34.28\%, and 34.32\%, respectively, with a maximum deviation of less than 0.11 pp (see Table~\ref{tab:llama31_8b_sample}). These results suggest that small sample sizes yield relatively stable and reliable results.

\section{Conclusion}
We introduce SocioBench, a cross-cultural benchmark using large-scale real-world sociological survey data to evaluate LLMs' ability to model human behavioral patterns. Through demographic role-play prompts, models generate answers that enable a systematic assessment of alignment with empirically observed social attitudes. 


\section*{Limitations}
\paragraph{Long-Term Data Sustainability.}
SocioBench relys on the static data of ISSP question–answer pairs and respondent answers. Although these data represent the currently newest survey round results, they cannot track longer-term attitudinal drift. 

\paragraph{Evaluation of Dynamism and Openness.}
The current evaluation relies solely on accuracy, focusing on matching answers at the individual level; and its evaluation of dynamism is insufficient. 



\section*{Ethic Statement}
The SocioBench dataset is based on ISSP\footnote{\url{https://www.issp.org}}. And we contacted the official data provider GESIS (Leibniz Institute for the Social Sciences; isspservice@gesis.org) via email and obtained explicit written permission authorizing the use of the dataset for this study and for publication. Use of the SocioBench must strictly adhere to the data usage requirements of the ISSP and GESIS\footnote{\url{https://www.gesis.org/en/institute/data-usage-terms}}.

\section*{Acknowledgements}
The research is supported by National Key R\&D Program of China (Grant No. 2023YFF1204800) and National Natural Science Foundation of China (Grant No. 62176058). And we sincerely thank the experts at GESIS for their crucial data support and guidance throughout this study.

\bibliography{anthology,custom}
\bibliographystyle{acl_natbib}

\newpage
\appendix
\onecolumn

\section{Details of Data Statistics}
\label{sec:data_details}

\subsection{Statistics and Analysis}
\label{sec:data_stat}

\begin{figure}[htbp]
  \centering
  \includegraphics[width=1\linewidth]{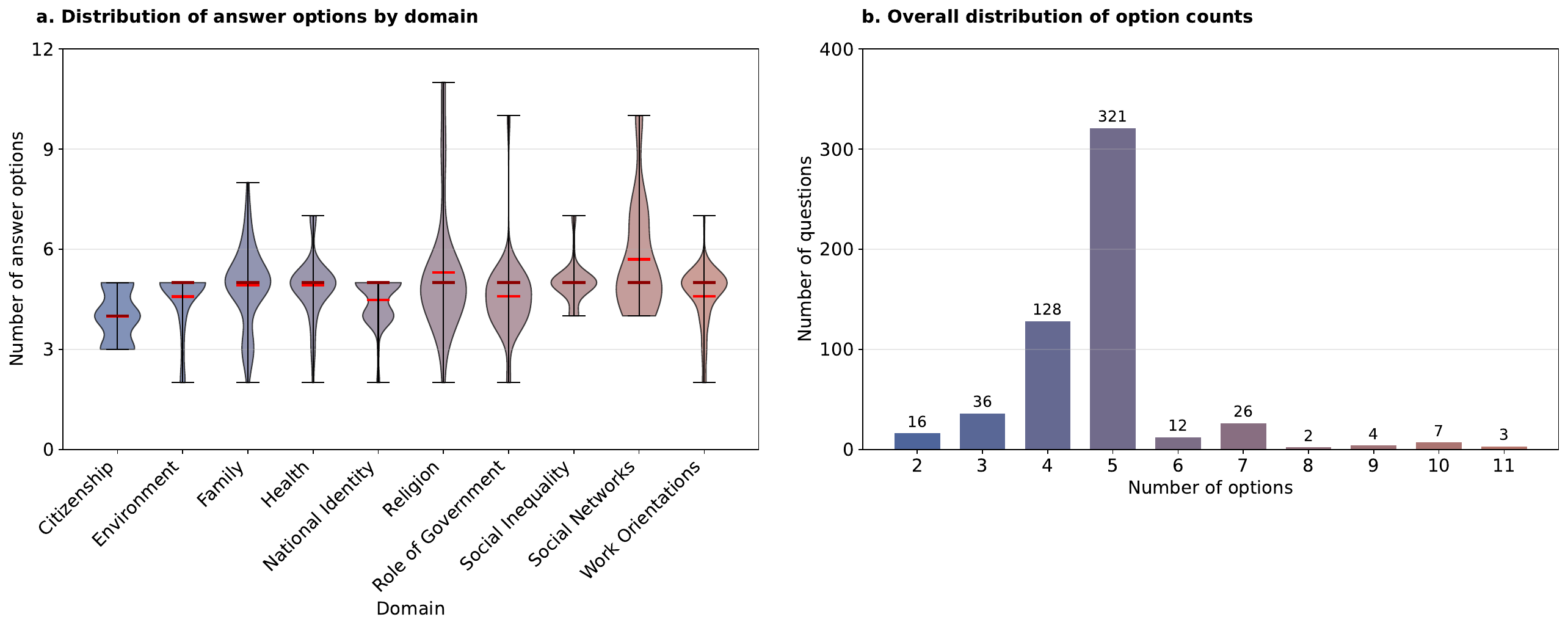}
  \caption{
    \textbf{Question and Answer Option Distribution Analysis across ISSP Survey Domains.}
    (a) shows the distribution of answer options per question across domains using violin plots. The width of each violin represents the density of questions with that number of options. The red line indicates the mean number of options, while the dark red line shows the median number of options for each domain. The black lines represent the data range (minimum to maximum values).
    (b) displays the overall distribution of questions grouped by answer option count across the entire dataset, showing how many questions have 2, 3, 4, 5, etc. answer options in total.
  }
  \label{fig:question_option_distribution}
\end{figure}

Figure~\ref{fig:question_option_distribution} provides a detailed overview of the structural characteristics of questionnaire items in the SocioBench dataset. 

\subsection{Demographic Information Distribution within the Citizenship Domain}
\label{sec:Demographic_Information}
Table~\ref{tab:citizenship_country_dist_multi} to Table~\ref{tab:citizenship_religious_dist_single} show the distribution statistics of some demographic information in the citizenship domain, including gender, country, year of birth, educational background, and religion.

\begin{table}[htbp]
\centering
\footnotesize
\setlength{\tabcolsep}{2pt}
\begin{minipage}{0.32\textwidth}
\centering
\begin{tabular}{lcc}
\toprule
Value & Freq. & Pct. \\
\midrule
Austria & 16 & 3.2\% \\
Taiwan, China & 16 & 3.2\% \\
Australia & 16 & 3.2\% \\
Croatia & 16 & 3.2\% \\
Chile & 16 & 3.2\% \\
Lithuania & 15 & 3.0\% \\
Belgium & 15 & 3.0\% \\
Netherlands & 15 & 3.0\% \\
Korea (South) & 15 & 3.0\% \\
Slovakia & 15 & 3.0\% \\
Turkey & 15 & 3.0\% \\
\bottomrule
\end{tabular}
\end{minipage}
\hfill\begin{minipage}{0.32\textwidth}
\centering
\begin{tabular}{lcc}
\toprule
Value & Freq. & Pct. \\
\midrule
Venezuela & 15 & 3.0\% \\
United States of America & 15 & 3.0\% \\
Czech Republic & 15 & 3.0\% \\
Germany & 15 & 3.0\% \\
Russia & 15 & 3.0\% \\
Great Britain & 15 & 3.0\% \\
Spain & 15 & 3.0\% \\
Poland & 15 & 3.0\% \\
Georgia & 15 & 3.0\% \\
South Africa & 15 & 3.0\% \\
Norway & 15 & 3.0\% \\
\bottomrule
\end{tabular}
\end{minipage}
\hfill\begin{minipage}{0.32\textwidth}
\centering
\begin{tabular}{lcc}
\toprule
Value & Freq. & Pct. \\
\midrule
France & 15 & 3.0\% \\
Japan & 15 & 3.0\% \\
Philippines & 15 & 3.0\% \\
Israel & 15 & 3.0\% \\
India & 15 & 3.0\% \\
Finland & 15 & 3.0\% \\
Switzerland & 15 & 3.0\% \\
Slovenia & 15 & 3.0\% \\
Iceland & 15 & 3.0\% \\
Denmark & 15 & 3.0\% \\
Sweden & 15 & 3.0\% \\
\bottomrule
\end{tabular}
\end{minipage}
\caption{Demographic Profile of Citizenship Domain: Country Distribution. Freq. refers to the frequency of occurrence, Pct. refers to the percentage}
\label{tab:citizenship_country_dist_multi}
\end{table}

\begin{table}[htbp]
\centering
\footnotesize
\setlength{\tabcolsep}{1pt}
\begin{minipage}{0.19\textwidth}
\centering
\begin{tabular}{lcc}
\toprule
Value & Freq. & Pct. \\
\midrule
1975 & 15 & 3.0\% \\
1962 & 14 & 2.8\% \\
1961 & 14 & 2.8\% \\
1949 & 13 & 2.6\% \\
1963 & 13 & 2.6\% \\
1965 & 13 & 2.6\% \\
1958 & 13 & 2.6\% \\
1976 & 12 & 2.4\% \\
1981 & 12 & 2.4\% \\
1967 & 12 & 2.4\% \\
1979 & 12 & 2.4\% \\
1964 & 12 & 2.4\% \\
1985 & 12 & 2.4\% \\
1969 & 12 & 2.4\% \\
\bottomrule
\end{tabular}
\end{minipage}
\hfill\begin{minipage}{0.19\textwidth}
\centering
\begin{tabular}{lcc}
\toprule
Value & Freq. & Pct. \\
\midrule
1960 & 12 & 2.4\% \\
1977 & 10 & 2.0\% \\
1951 & 10 & 2.0\% \\
1971 & 10 & 2.0\% \\
1952 & 10 & 2.0\% \\
1992 & 10 & 2.0\% \\
1970 & 9 & 1.8\% \\
1938 & 9 & 1.8\% \\
1974 & 9 & 1.8\% \\
1982 & 9 & 1.8\% \\
1942 & 9 & 1.8\% \\
1950 & 9 & 1.8\% \\
1989 & 8 & 1.6\% \\
1983 & 8 & 1.6\% \\
\bottomrule
\end{tabular}
\end{minipage}
\hfill\begin{minipage}{0.19\textwidth}
\centering
\begin{tabular}{lcc}
\toprule
Value & Freq. & Pct. \\
\midrule
1972 & 8 & 1.6\% \\
1953 & 8 & 1.6\% \\
1954 & 8 & 1.6\% \\
1955 & 8 & 1.6\% \\
1994 & 7 & 1.4\% \\
1947 & 7 & 1.4\% \\
1980 & 7 & 1.4\% \\
1984 & 7 & 1.4\% \\
1973 & 7 & 1.4\% \\
1944 & 7 & 1.4\% \\
1957 & 6 & 1.2\% \\
1986 & 6 & 1.2\% \\
1956 & 6 & 1.2\% \\
1978 & 6 & 1.2\% \\
\bottomrule
\end{tabular}
\end{minipage}
\hfill\begin{minipage}{0.19\textwidth}
\centering
\begin{tabular}{lcc}
\toprule
Value & Freq. & Pct. \\
\midrule
1943 & 6 & 1.2\% \\
1988 & 6 & 1.2\% \\
1946 & 5 & 1.0\% \\
1933 & 5 & 1.0\% \\
1939 & 5 & 1.0\% \\
1948 & 5 & 1.0\% \\
1968 & 5 & 1.0\% \\
1993 & 5 & 1.0\% \\
1987 & 5 & 1.0\% \\
1966 & 4 & 0.8\% \\
1935 & 4 & 0.8\% \\
1959 & 4 & 0.8\% \\
1995 & 4 & 0.8\% \\
1937 & 3 & 0.6\% \\
\bottomrule
\end{tabular}
\end{minipage}
\hfill\begin{minipage}{0.19\textwidth}
\centering
\begin{tabular}{lcc}
\toprule
Value & Freq. & Pct. \\
\midrule
1934 & 3 & 0.6\% \\
1991 & 3 & 0.6\% \\
1936 & 3 & 0.6\% \\
1990 & 3 & 0.6\% \\
1941 & 2 & 0.4\% \\
1998 & 2 & 0.4\% \\
1932 & 2 & 0.4\% \\
1940 & 1 & 0.2\% \\
No answer & 1 & 0.2\% \\
1931 & 1 & 0.2\% \\
1925 & 1 & 0.2\% \\
1997 & 1 & 0.2\% \\
1996 & 1 & 0.2\% \\
1945 & 1 & 0.2\% \\
\bottomrule
\end{tabular}
\end{minipage}
\caption{Demographic Profile of Citizenship Domain: Birth Year Distribution}
\label{tab:citizenship_birth_year_dist_5col}
\end{table}

\begin{table}[htbp]
\centering
\small
\setlength{\tabcolsep}{5pt}
\begin{tabular*}{\textwidth}{@{\extracolsep{\fill}}lcc}
\toprule
Value & Freq. & Pct. \\
\midrule
Upper secondary (programs that allow entry to university) & 122 & 24.4\% \\
Lower level tertiary, first stage (also technical schools at a tertiary level) & 111 & 22.2\% \\
Lower secondary (secondary completed does not allow entry to university: obligatory school) & 106 & 21.2\% \\
Upper level tertiary (Master, Doctor) & 65 & 13.0\% \\
Post secondary, non-tertiary (other upper secondary programs toward labour market or technical formation) & 59 & 11.8\% \\
Primary school (elementary education) & 22 & 4.4\% \\
No formal education & 14 & 2.8\% \\
No answer & 1 & 0.2\% \\
\bottomrule
\end{tabular*}
\caption{Demographic Profile of Citizenship Domain: Education Level Distribution}
\label{tab:citizenship_education_level_dist_full_width}
\end{table}

\begin{table}[htbp]
\centering
\footnotesize
\setlength{\tabcolsep}{3pt}
\begin{tabular*}{0.4\textwidth}{@{\extracolsep{\fill}}lcc}
\toprule
Value & Freq. & Pct. \\
\midrule
Male & 257 & 51.4\% \\
Female & 243 & 48.6\% \\
\bottomrule
\end{tabular*}
\caption{Demographic Profile of Citizenship Domain: Gender Distribution}
\label{tab:citizenship_gender_dist_compact}
\end{table}

\begin{table}[htbp]
\centering
\footnotesize
\setlength{\tabcolsep}{3pt}
\begin{tabular*}{0.4\textwidth}{@{\extracolsep{\fill}}lcc}
\toprule
Value & Freq. & Pct. \\
\midrule
No religion             & 140 & 28.0\% \\
Catholic                & 139 & 27.8\% \\
Protestant              & 100 & 20.0\% \\
Orthodox                &  26 &  5.2\% \\
Islamic                 &  25 &  5.0\% \\
Other Christian         &  17 &  3.4\% \\
Buddhist                &  14 &  2.8\% \\
Hindu                   &  14 &  2.8\% \\
Jewish                  &  10 &  2.0\% \\
Other Asian Religions   &   5 &  1.0\% \\
No answer               &   3 &  0.6\% \\
Other Religions         &   3 &  0.6\% \\
Refused                 &   3 &  0.6\% \\
Information insufficient&   1 &  0.2\% \\
\bottomrule
\end{tabular*}
\caption{Demographic Profile of Citizenship Domain: Religious Affiliation Distribution}
\label{tab:citizenship_religious_dist_single}
\end{table}

\clearpage
\subsection{Data example}
\label{sec:Data_example}

\begin{figure}[htbp]
  \centering
  \includegraphics[width=0.95\linewidth]{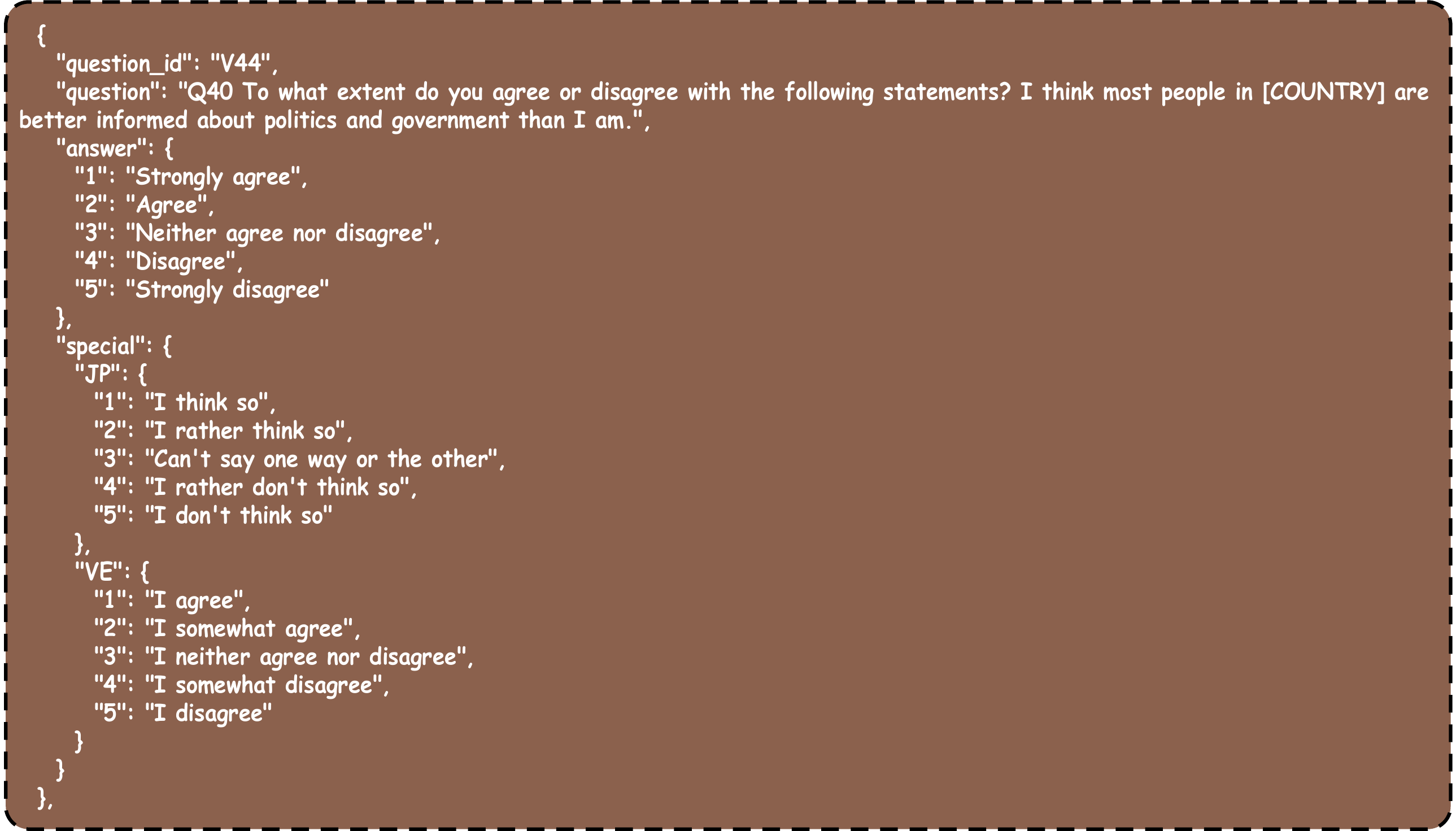}
  \caption{SocioBench Dataset: Questions and answers in social survey questionnaires}
  \label{fig:Social_survey_scale_respondent_profile_label}
\end{figure}

\begin{figure}[htbp]
  \centering
  \includegraphics[width=0.95\linewidth]{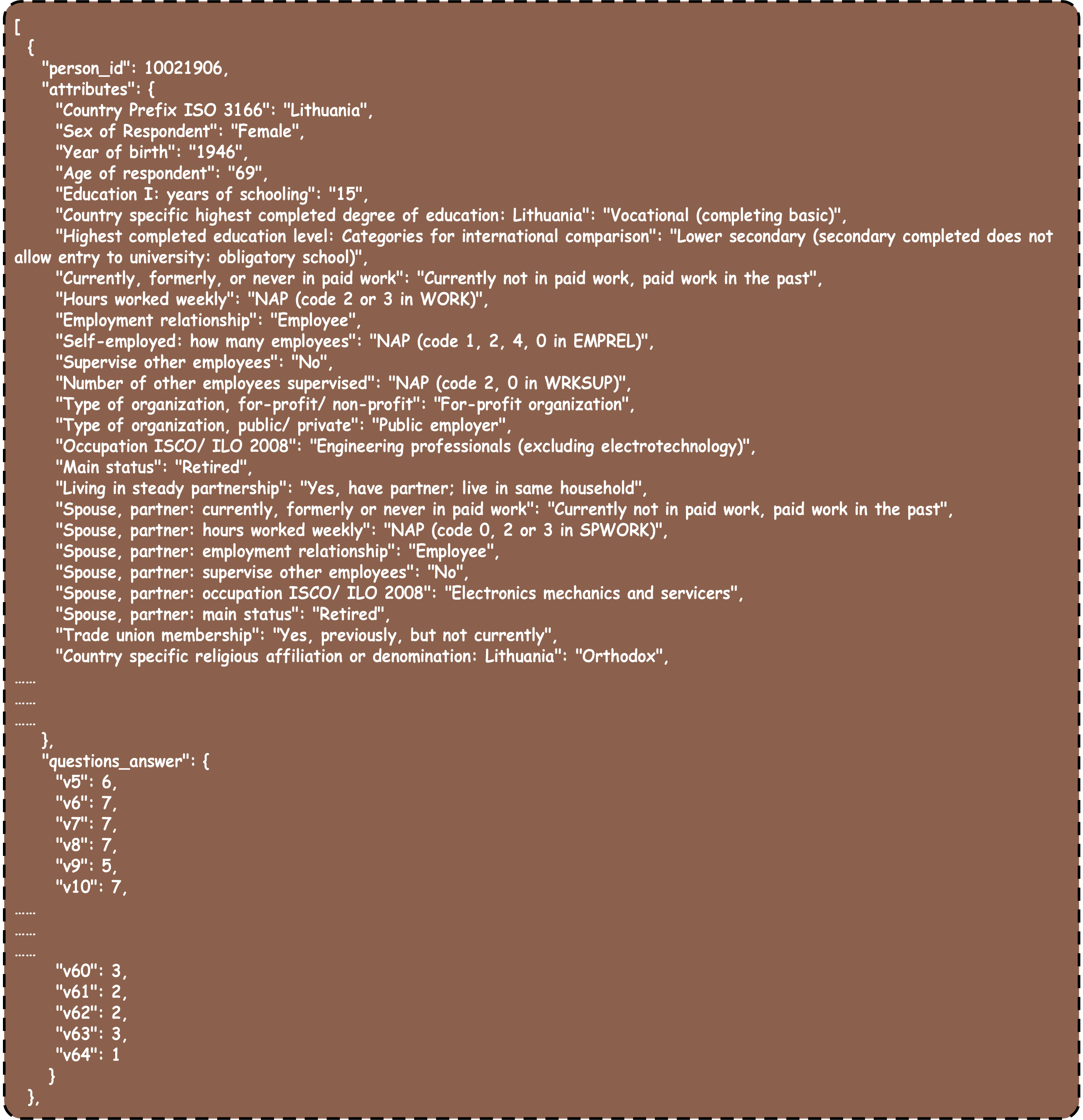}
  \caption{SocioBench Dataset: respondent demographic information and Ground-truth answers}
  \label{fig:Ground-truth_respondent_demographic_information}
\end{figure}

Figure~\ref{fig:Social_survey_scale_respondent_profile_label} and Figure~\ref{fig:Ground-truth_respondent_demographic_information} respectively show the questionnaire data, respondent profile data and ground-truth answer data contained in the SocioBench dataset. Figure~\ref{fig:Social_survey_scale_respondent_profile_label} shows the Q\&A data processing for special countries. For example, for question V44, when the respondent's country code is equal to the country code of "special" in the dataset, the corresponding question option in "special" replaces the question option in "answer" and asks the question.

\clearpage
\section{Data Curation Details}
\label{sec:data_curation}

\begin{figure}[!htbp]
    \centering
    \includegraphics[width=0.95\textwidth]{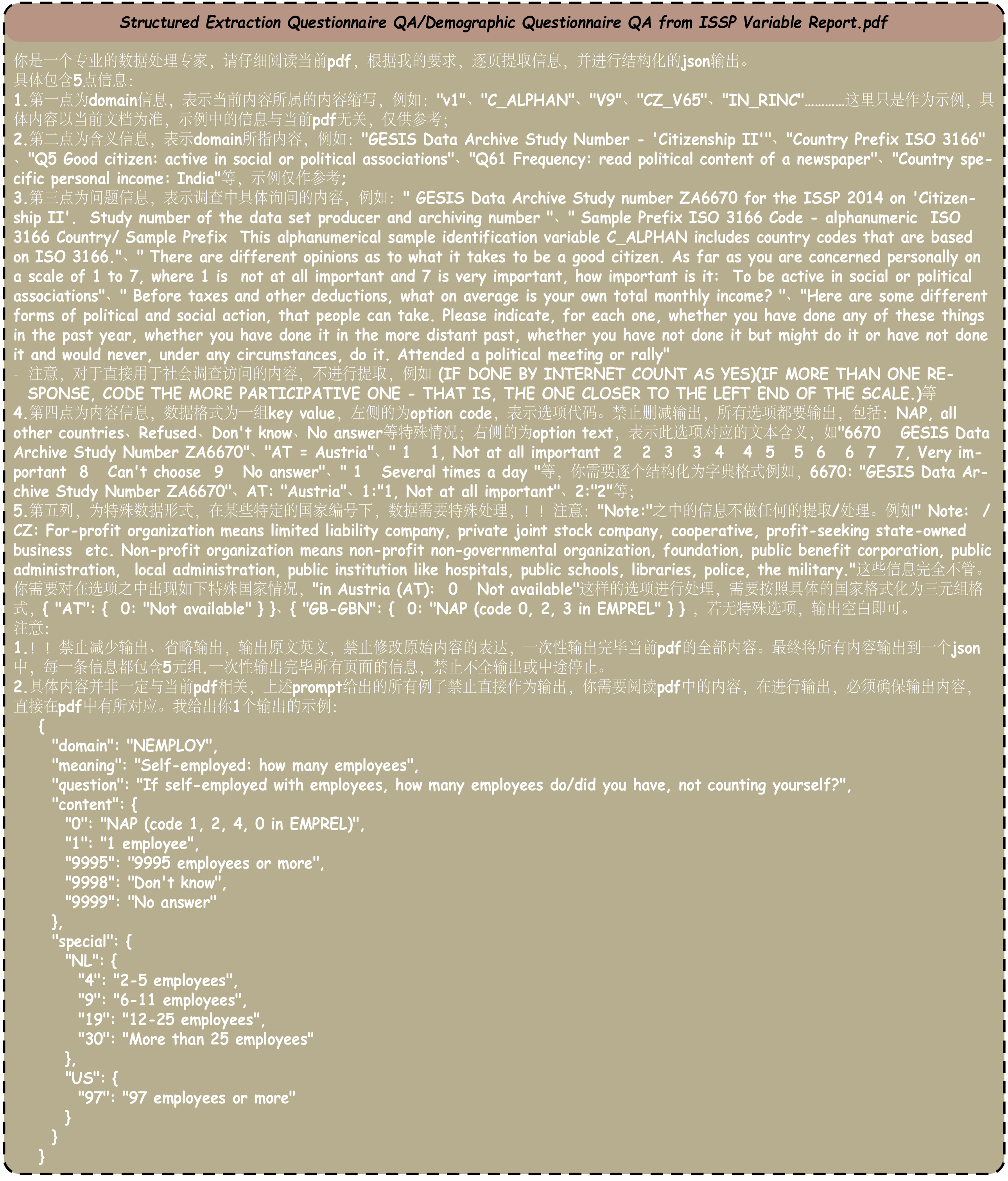}
    \caption{Structured Extraction Questionnaire QA/Demographic Questionnaire QA from ISSP Variable Report.pdf (Chinese)}
    \label{Structured Extraction Questionnaire QA/Demographic Questionnaire QA from ISSP Variable Report.pdf (Chinese)}
\end{figure}
\begin{figure}[t]
    \centering
    \includegraphics[width=0.95\textwidth]{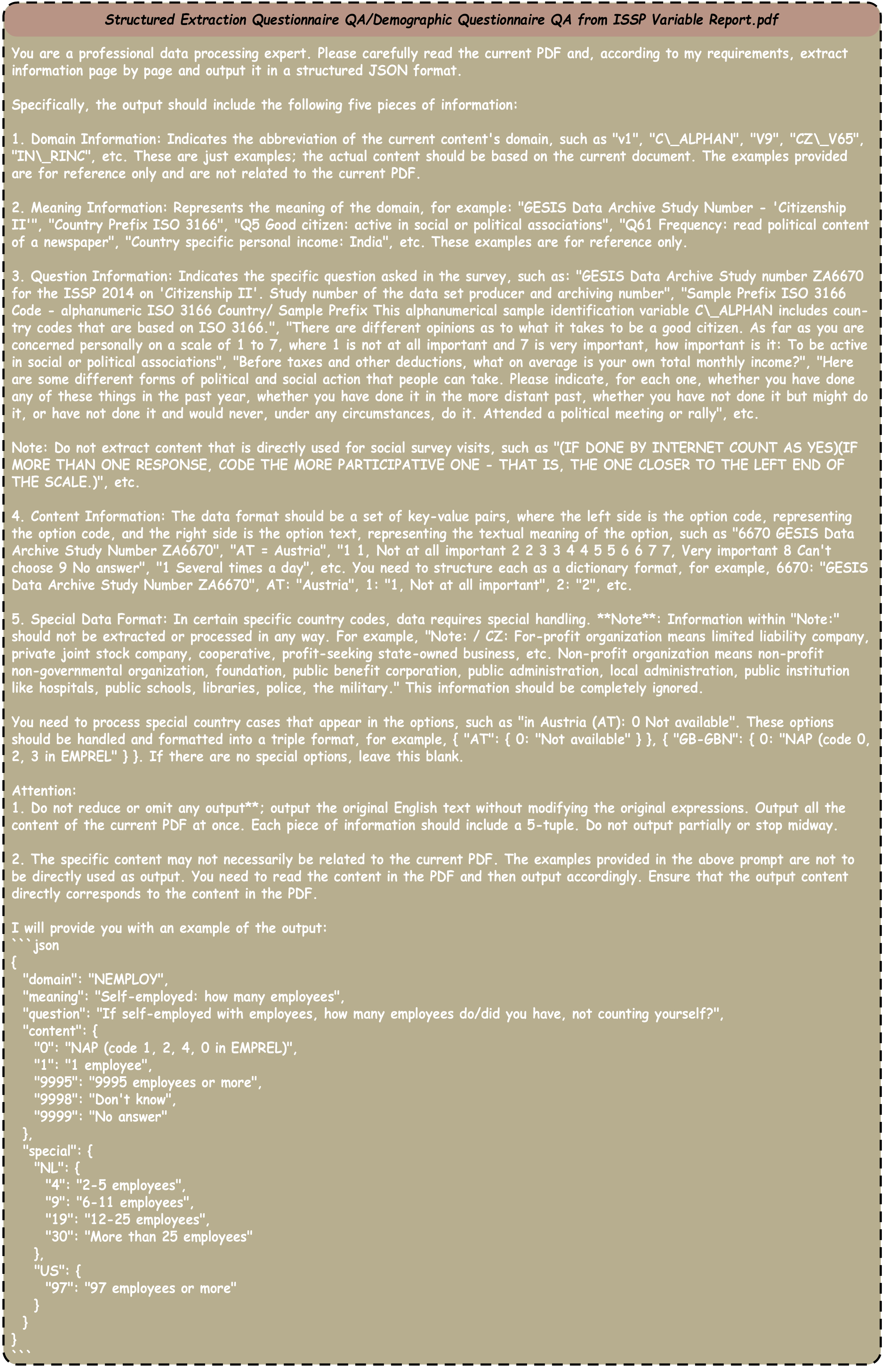}
    \caption{Structured Extraction Questionnaire QA/Demographic Questionnaire QA from ISSP Variable Report.pdf (English)}
    \label{Structured Extraction Questionnaire QA/Demographic Questionnaire QA from ISSP Variable Report.pdf (English)}
\end{figure}

Figures~\ref{Structured Extraction Questionnaire QA/Demographic Questionnaire QA from ISSP Variable Report.pdf (Chinese)} and~\ref{Structured Extraction Questionnaire QA/Demographic Questionnaire QA from ISSP Variable Report.pdf (English)} show structured Questionnaire QA/Demographic Questionnaire QA examples extracted from the ISSP Variable Report.pdf, in Chinese and English versions, respectively.

\clearpage
\section{LLMs Role-playing Prompt template}
\label{sec:LLMs Role-playing Prompt template}
\begin{figure}[!htbp]
    \centering
    \includegraphics[width=0.95\textwidth]{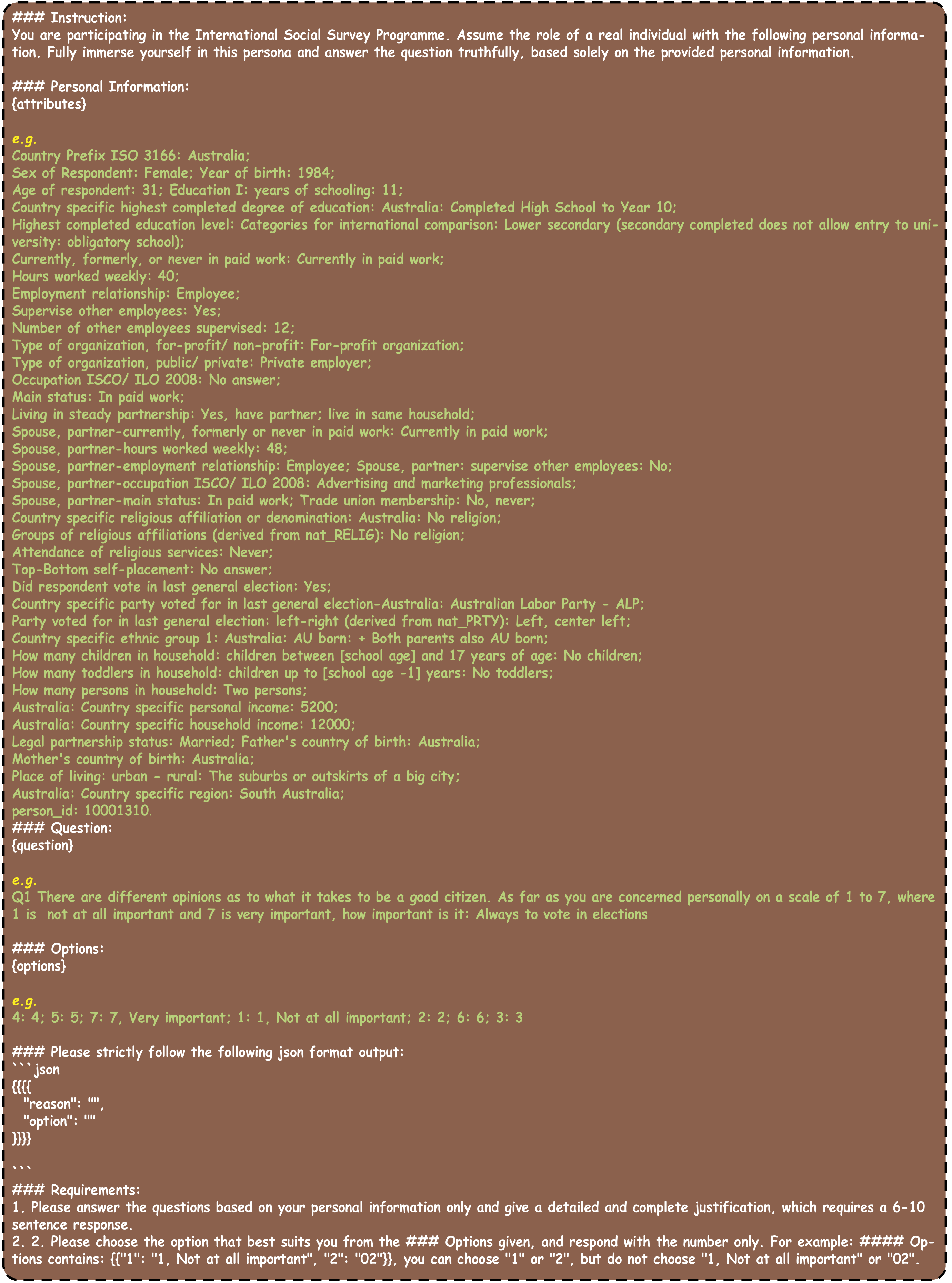}
    \caption{Prompt Template for LLMs Role-playing Respondents in Social Survey Scenarios}
    \label{sec:Social Survey Respondents RPLAs Prompt Word Template}
\end{figure}

Figure~\ref{sec:Social Survey Respondents RPLAs Prompt Word Template} shows the prompt templates passed to the LLMs along with examples.

\clearpage
\section{Comparison of Qwen3-32B With and Without "Think" Mode}
\label{sec:Comparison of Qwen3-32B With and Without Think Mode}
\begin{figure}[!htbp]
    \centering
    \includegraphics[width=0.95\textwidth]{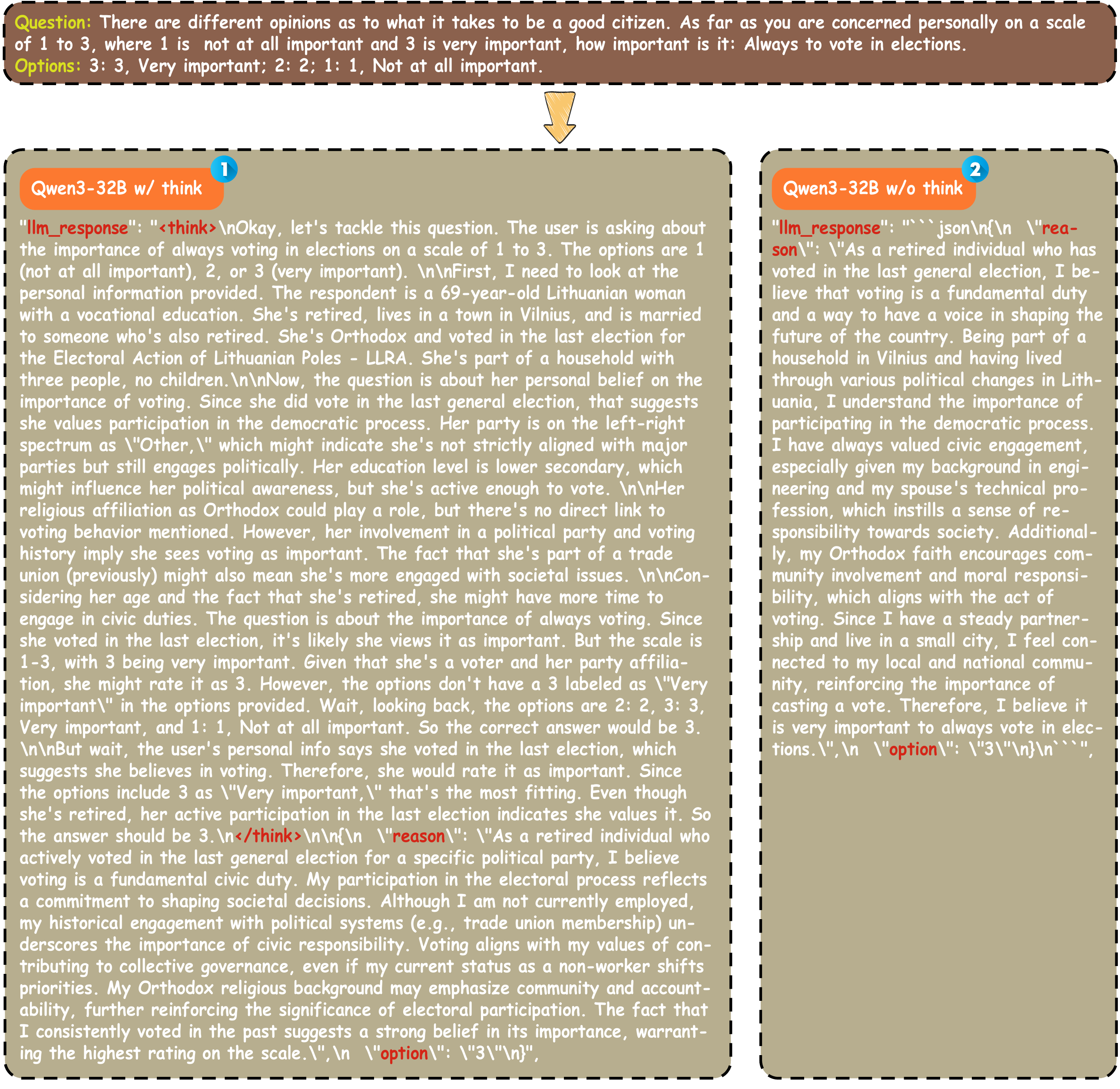}
    \caption{Qwen3-32B response comparison on the same question}
    \label{Qwen3-32B response comparison on the same question-think}
\end{figure}

Figure~\ref{Qwen3-32B response comparison on the same question-think} shows the Qwen3-32B response comparison on the same question.

\clearpage
\section{Comparison of Qwen3-32B With and Without "Reason" in prompt}
\label{sec:Comparison of Qwen3-32B With and Without Reason Mode}
\begin{figure}[!htbp]
    \centering
    \includegraphics[width=0.95\textwidth]{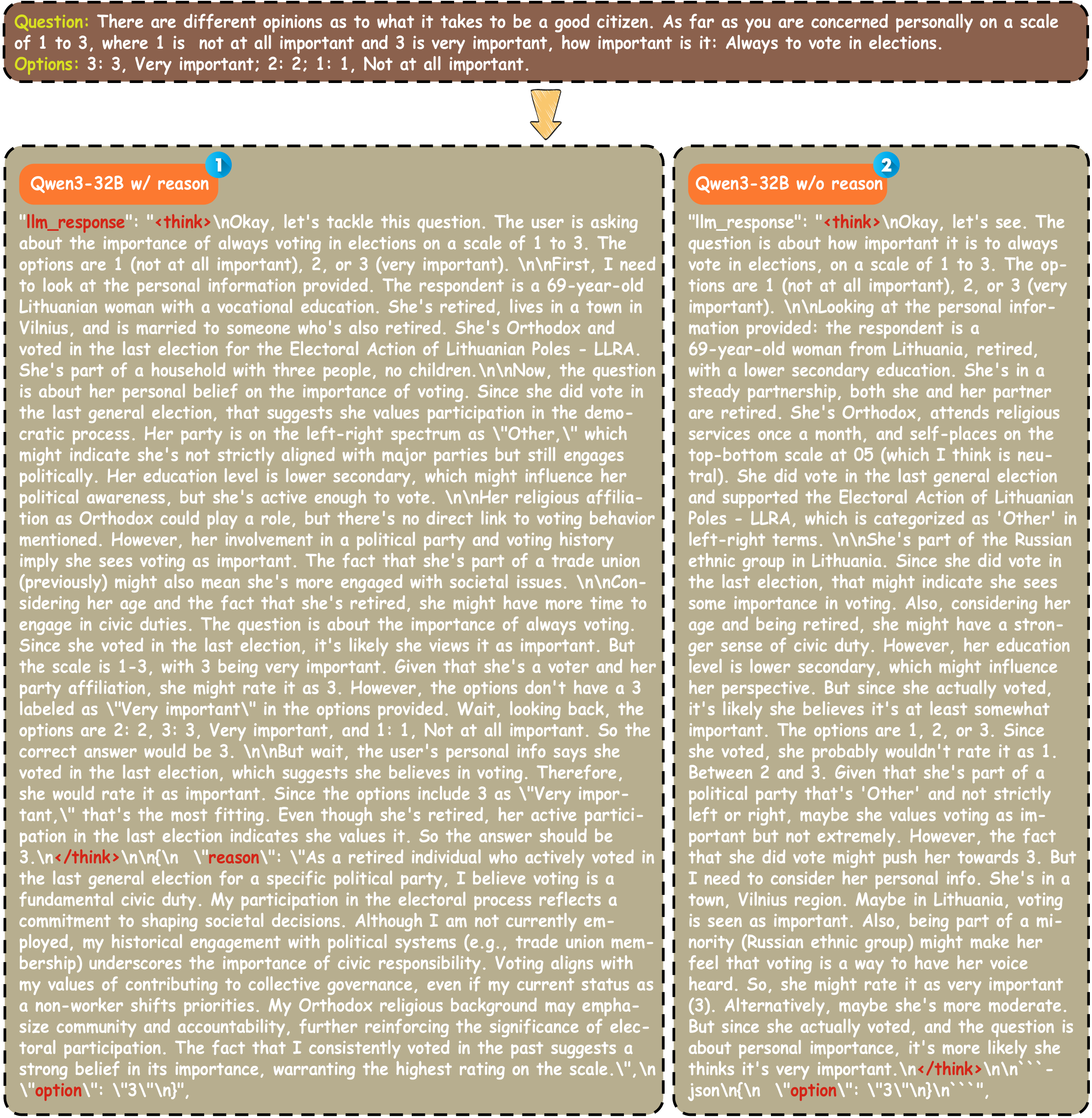}
    \caption{Qwen3-32B response comparison on the same question}
    \label{Qwen3-32B response comparison on the same question-reason}
\end{figure}

Figure~\ref{Qwen3-32B response comparison on the same question-reason} shows the Qwen3-32B response comparison on the same question.

\clearpage
\section{Example comparison of the option distribution for real respondents and LLM-generated responses.}
\label{sec:Example-comparison}

Focusing on the \textit{Family} and \textit{Health and Health Care} domains, we conducted a further analysis comparing real respondents with four representative models—Qwen2.5-72B-Instruct, Qwen3-32B, Llama-3.3-70B-Instruct, and GPT-4o—by sampling ten questions and examining the response-option distributions. 

As shown in Figure~\ref{fig:family_distribution_comparison} and Figure~\ref{fig:health_distribution_comparison}, although the ground-truth results exhibit skewed distributions (i.e., options are concentrated in several categories), the LLM-generated responses make this skewness more pronounced, with Llama-3.3-70B-Instruct showing the most marked concentration. Conversely, we observe that Qwen3-32B tends to produce more uniform option distributions.

\begin{figure}[h!]
    \centering
    \includegraphics[
        page=1,
        width=0.92\textwidth
    ]{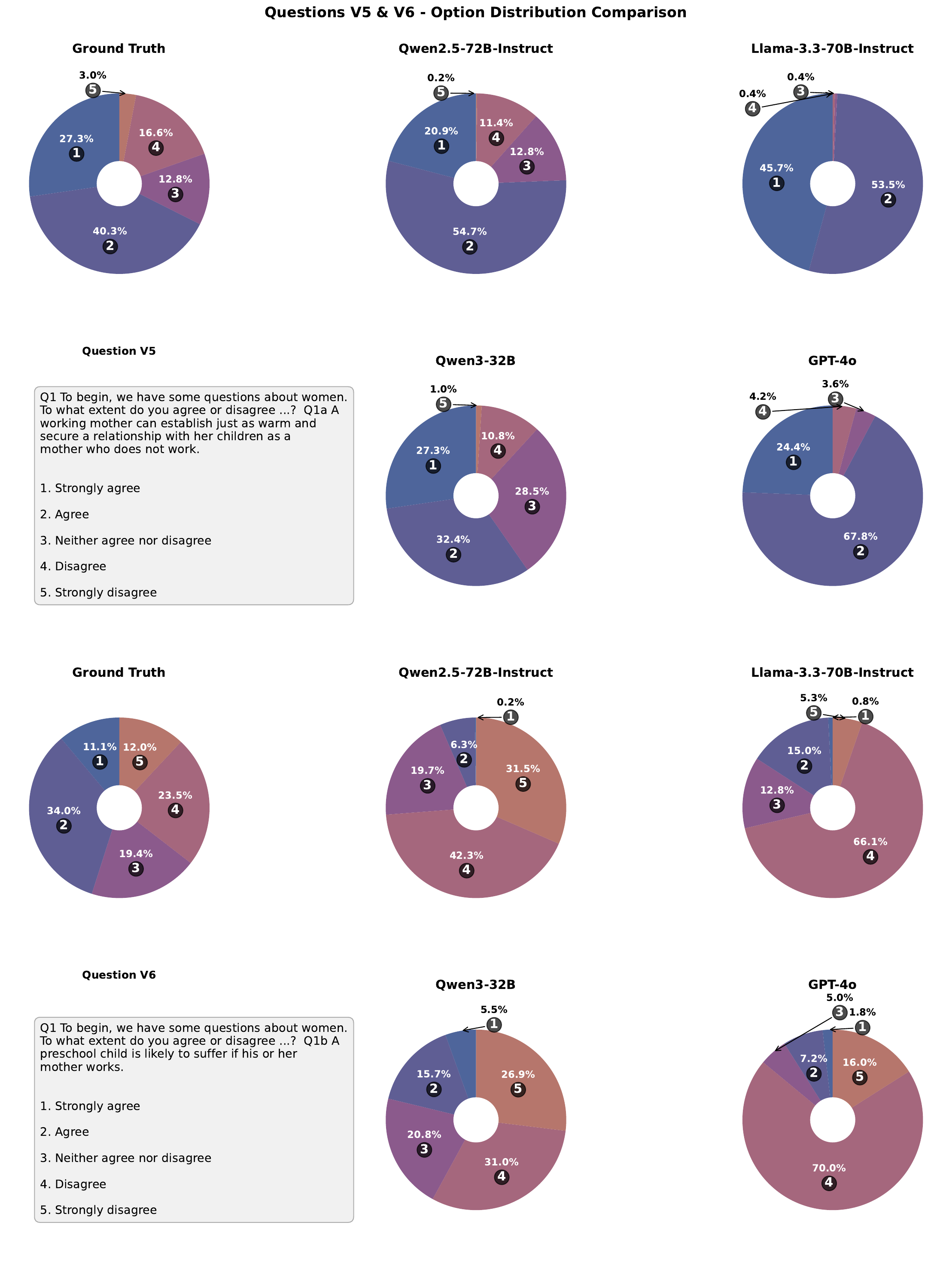}
\end{figure}

\includepdf[
    pages=2,
    width=0.95\textwidth,
    pagecommand={\thispagestyle{empty}}
]{figures/Family_distribution_comparison.pdf}

\includepdf[
    pages=3,
    width=0.95\textwidth,
    pagecommand={
        \thispagestyle{empty}
        \vspace*{\fill}
        \captionsetup{justification=centering}
        \captionof{figure}{Comparison of the option distribution in the family domain}
        \label{fig:family_distribution_comparison}
    }
]{figures/Family_distribution_comparison.pdf}

\includepdf[
pages=1,
width=0.95\textwidth,
pagecommand={\thispagestyle{empty}}
]{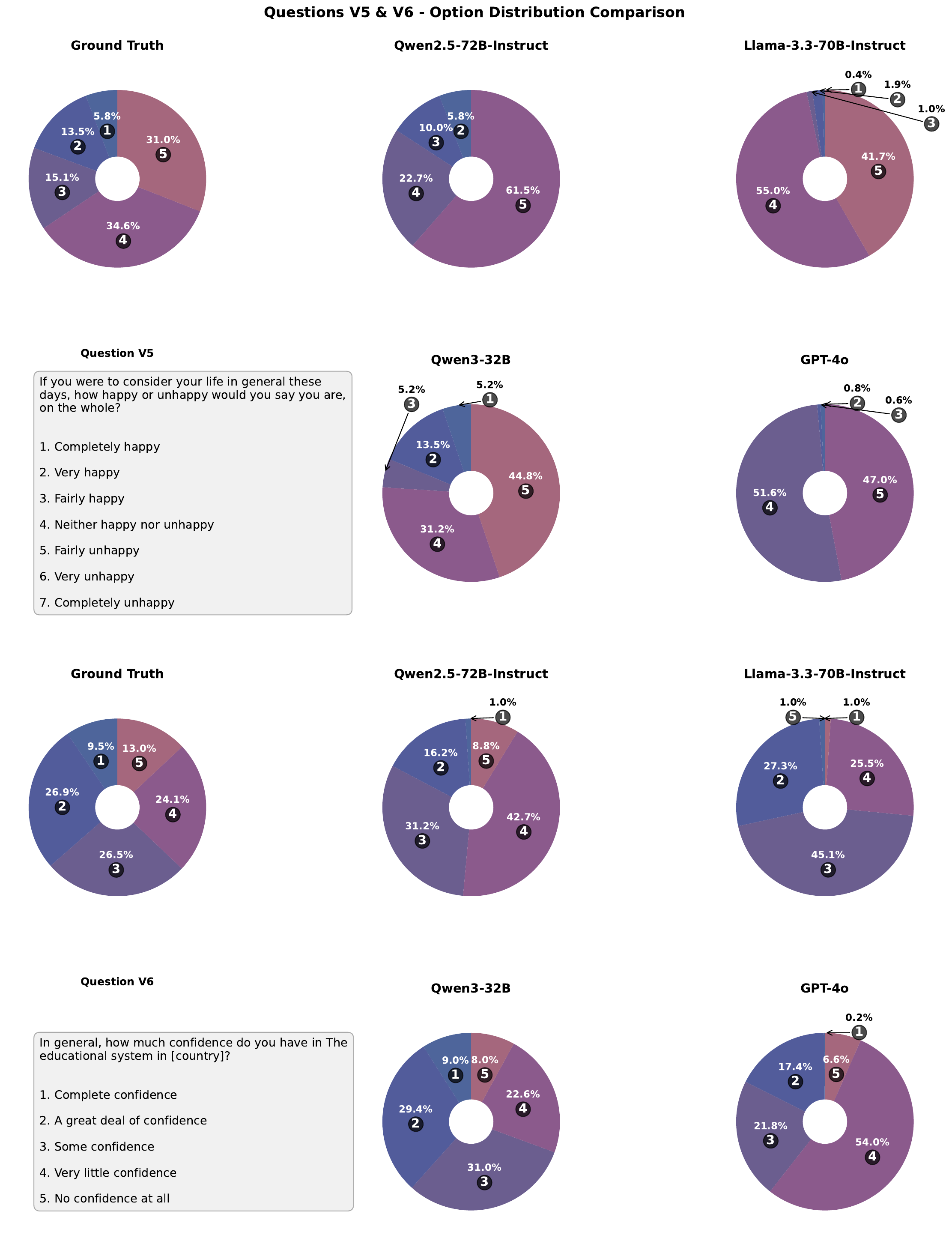}

\includepdf[
pages=2,
width=0.95\textwidth,
pagecommand={\thispagestyle{empty}}
]{figures/Health_distribution_comparison.pdf}

\includepdf[
pages=3,
width=0.95\textwidth,
pagecommand={
\thispagestyle{empty}
\vspace*{\fill}
\captionsetup{justification=centering}
\captionof{figure}{Comparison of the option distribution in the health domain}
\label{fig:health_distribution_comparison}
}
]{figures/Health_distribution_comparison.pdf}

\clearpage
\section{Supplementary experimental results and findings}
\label{sec:Supplementary experimental results and findings}

    \paragraph{Data Sampling Ratios Comparison.} 
    For the experimental result of dataset sampling ratios, please refer to the Table~\ref{tab:llama31_8b_sample}.

    \begin{table}[htbp]   
      \centering
      \TableFont                     
      \setlength{\tabcolsep}{3pt}
      \begin{tabular*}{.6\linewidth}{@{\extracolsep{\fill}}lccc}
        \toprule
              & n = 5000 & n = 10000 & n = 20000 \\
        \midrule
        Citizen   & 40.43 & 40.15 & 40.07 \\
        Enviro    & 32.11 & 32.33 & 32.08 \\
        Family    & 31.89 & 32.82 & 33.19 \\
        Health    & 32.21 & 32.55 & 32.47 \\
        Nat.Ident & 33.37 & 33.37 & 33.27 \\
        Religion  & 36.99 & 37.17 & 36.83 \\
        R.Gov     & 35.27 & 34.85 & 34.87 \\
        S.Ineq    & 31.47 & 31.03 & 31.48 \\
        S.Net     & 34.99 & 34.43 & 35.02 \\
        Work      & 33.39 & 34.10 & 33.94 \\
        \midrule
        Avg.      & 34.21 & 34.28 & 34.32 \\
        \bottomrule
      \end{tabular*}
      \vspace{-2mm}
      \caption{Results of Llama-3.1-8B-Instruct under different sampling ratios (n denotes the number of respondents under 10 domains).} 
      \label{tab:llama31_8b_sample}
      \vspace{-4mm}
    \end{table}

    \paragraph{How does requiring a reason in responses  affect LLMs' behavioral simulation?} 
    To analyze how providing reasons impacts the evaluation, we conducted experiments on Qwen3-8B and Qwen3-32B, comparing two response strategies: \textit{Option-only} vs. \textit{Reason \& Option}. The results indicate that including reasons has a minor effect on performance. In fact, it leads to a slight decrease in accuracy, as detailed in Table~\ref{tab:qwen3_reason}. We analyse that this may be due to the cognitive overhead or response biases, which can interfere with the model's intrinsic decision-making process. An output example can be found in Appendix~\ref{sec:Comparison of Qwen3-32B With and Without Reason Mode}.
    
    \begin{table}[htbp]   
      \centering
      \TableFont                     
      \setlength{\tabcolsep}{3pt}
      \begin{tabular*}{.6\linewidth}{@{\extracolsep{\fill}}lcccc}
        \toprule
              & 8B w/\;R & B w/o\;R & 32B w/\;R & 32B w/o\;R \\
        \midrule
        Citizen      & 40.28 & 39.96 & 43.60 & 44.18 \\
        Enviro       & 32.70 & 32.64 & 34.12 & 34.78 \\
        Family       & 33.07 & 33.61 & 34.53 & 35.37 \\
        Health       & 33.98 & 34.79 & 33.53 & 34.21 \\
        Nat.Ident    & 33.12 & 34.45 & 32.64 & 34.79 \\
        Religion     & 37.58 & 37.62 & 38.90 & 39.52 \\
        R.Gov        & 34.65 & 34.50 & 35.52 & 35.89 \\
        S.Ineq       & 30.83 & 30.81 & 33.16 & 33.14 \\
        S.Net        & 34.38 & 34.71 & 35.31 & 36.54 \\
        Work         & 34.20 & 35.55 & 35.25 & 35.18 \\
        \midrule
        Avg.         & 34.48 & \textbf{34.87} & 35.66 & \textbf{36.36} \\
        \bottomrule
      \end{tabular*}
      \vspace{-2mm}
      \caption{Results of Qwen3 models with/without reason in response (R indicates the reason why the LLM selected this option when responding.}
      \label{tab:qwen3_reason}
      \vspace{-4mm}
    \end{table}

    \paragraph{How thinking modes shape LLMs' behavioral simulation?}
    For the experimental result of how thinking and reasoning processes affect behavioral simulation in social survey scenarios, please refer to the Table~\ref{tab:qwen3_think}.
    
    \begin{table}[htbp]
      \centering
      \TableFont                     
      \setlength{\tabcolsep}{3pt}
      \begin{tabular*}{.6\linewidth}{@{\extracolsep{\fill}}lcccc}
        \toprule
              & 8B w/\;T & 8B w/o\;T & 32B w/\;T & 32B w/o\;T \\
        \midrule
        Citizen      & 40.28 & 42.34 & 43.60 & 43.52 \\
        Enviro       & 32.70 & 32.66 & 34.12 & 32.63 \\
        Family       & 33.07 & 30.36 & 34.53 & 32.05 \\
        Health       & 33.98 & 32.23 & 33.53 & 33.52 \\
        Nat.Ident    & 33.12 & 33.59 & 32.64 & 31.86 \\
        Religion     & 37.58 & 37.52 & 38.90 & 37.90 \\
        R.Gov        & 34.65 & 32.94 & 35.52 & 35.31 \\
        S.Ineq       & 30.83 & 30.78 & 33.16 & 32.15 \\
        S.Net        & 34.38 & 33.03 & 35.31 & 35.52 \\
        Work         & 34.20 & 34.25 & 35.25 & 33.27 \\
        \midrule
        Avg.         & \textbf{34.48} & 33.97 & \textbf{35.66} & 34.77 \\
        \bottomrule
      \end{tabular*}
      \vspace{-2mm}
      \caption{Results of Qwen3 Models With/Without Think Mode (T denotes the think mode; 8B and 32B denote Qwen3--8B and Qwen3--32B, respectively).}
      \label{tab:qwen3_think}
      \vspace{-4mm}
    \end{table}

    \paragraph{Comparison Across Survey Rounds.}

    Because the ISSP determines its annual sociological topics through general meetings and typically fields one survey per domain each year, we conducted additional, extensive experiments to compare how survey rounds from different years within the same domain affect benchmark results. Using Llama-3.3-70B-Instruct, we performed experiments for \textit{Environment}, \textit{Health and Healthcare}, \textit{National Identity,} \textit{Religion}, \textit{Role of Government}, \textit{Social Inequality}, and \textit{Work Orientations}. By contrast, for \textit{Citizenship}, \textit{Family and Changing Gender Roles}, and \textit{Social Networks}, limitations imposed by the data format of the Variable Reports files prevented us from extracting fully structured datasets; therefore, we did not carry out further experiments on these domains, see Table~\ref{tab:Comparison of Benchmark Accuracy Across Different survey rounds}.

    Across the seven domains with two waves, temporal changes remain modest and bidirectional: \textit{Religion} (+1.95 pp, 2008→2018), \textit{Role of Government} (+1.28 pp, 2006→2016), and \textit{Environment} (+1.28 pp, 2010→2020) show small improvements, while \textit{Work Orientations} (-4.90 pp, 2005→2015), \textit{Health and Healthcare} (-2.48 pp, 2011→2021), \textit{National Identity} (-1.22 pp, 2003→2013), and \textit{Social Inequality} (-1.36 pp, 2009→2019) decline. Averaged across these pairs, the later wave's accuracy is slightly lower by 0.78 pp than the earlier one (37.90\% vs. 38.68\%), indicating no systematic drift over time.
    
    The benchmark (bold) years used in SocioBench yield an average accuracy of 37.90\% (SD=1.90; range 35.73–41.26). The strongest results occur in \textit{Religion} (41.26\%) and \textit{Role of Government} (39.19\%). A similar pattern is observed in the earlier, non-benchmark waves, which exhibit a comparable mean accuracy of 38.68\% (SD=2.54; range 34.69–43.70), with \textit{Work Orientations} (43.70\%) and \textit{National Identity} (39.41\%) as the top performers. While temporal deltas show some variation—with \textit{Work Orientations} decreasing by 4.90 pp and \textit{Religion} increasing by 1.95 pp over their respective decade spans—most changes remain minor. This suggests that performance is driven more by domain-specific structure than by survey rounds.
    
\begin{figure}[!htbp]
        \centering
        \includegraphics[width=1\textwidth]{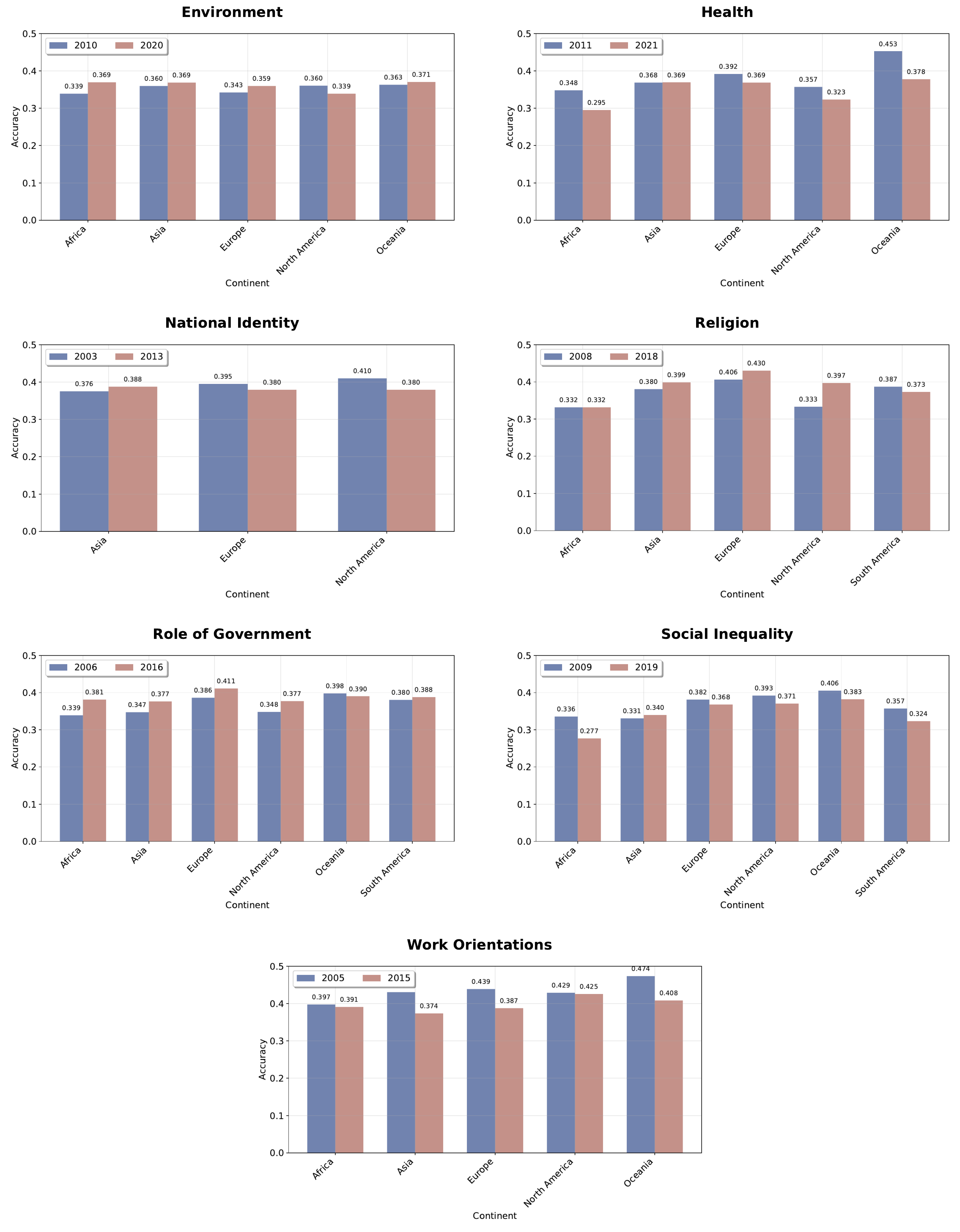}
        \caption{Comparison of benchmark accuracy across different continents in the two survey rounds.}
        \label{fig: Comparison of benchmark accuracy across different continents in the two survey rounds.}
    \end{figure}

    As observed from Figure~\ref{fig: Comparison of benchmark accuracy across different continents in the two survey rounds.}, within the same domain, the accuracy between the two survey rounds is highly consistent across continents. For instance, in the \textit{Environment}, performance in the first round is uniformly lower than in the second round for all continents. Conversely, in the \textit{Health and Healthcare} domain, the first round consistently outperforms the second across all continents. This indicates that while accuracy is influenced by the domain and the specific survey round, the benchmark performance demonstrates coordination and consistency across different continents.
    
    \begin{table}[htbp]
    \centering
    \TableFont                     
    \setlength{\tabcolsep}{3pt}
    \begin{tabular*}{.6\linewidth}{@{\extracolsep{\fill}}lcc}
    \toprule
    Domain & Year & Accuracy \\
    \midrule
    
    
    \multirow{2}{*}{Enviro}  & 2010 & 34.69 \\
                           & \textbf{2020} & 35.97 \\
    \midrule
    
    
    \multirow{2}{*}{Health}  & 2011 & 38.64 \\
                           & \textbf{2021} & 36.16 \\
    \midrule
    
    \multirow{2}{*}{Nat.Ident} & 2003 & 39.41 \\
                             & \textbf{2013} & 38.19 \\
    \midrule
    
    \multirow{2}{*}{Religion} & 2008 & 39.31 \\
                            & \textbf{2018} & 41.26 \\
    \midrule
    
    \multirow{2}{*}{R.Gov}    & 2006 & 37.91 \\
                            & \textbf{2016} & 39.19 \\
    \midrule
    
    \multirow{2}{*}{S.Ineq}   & 2009 & 37.09 \\
                            & \textbf{2019} & 35.73 \\
    \midrule
    
    
    \multirow{2}{*}{Work}     & 2005 & 43.70 \\
                            & \textbf{2015} & 38.80 \\
    \bottomrule
    \end{tabular*}
    \caption{Comparison of benchmark accuracy across survey rounds. Years set in bold correspond to the data years used in the SocioBench dataset, whereas years in regular (non-bold) type denote supplementary comparison waves.}
    \label{tab:Comparison of Benchmark Accuracy Across Different survey rounds}
    \end{table}

    \clearpage

\clearpage
\section{Subgroup analysis: Biases Across Demographic Information}
\label{sec:Subgroup_analysis}
\begin{figure*}[htbp]
        \centering
        \includegraphics[width=0.95\textwidth]{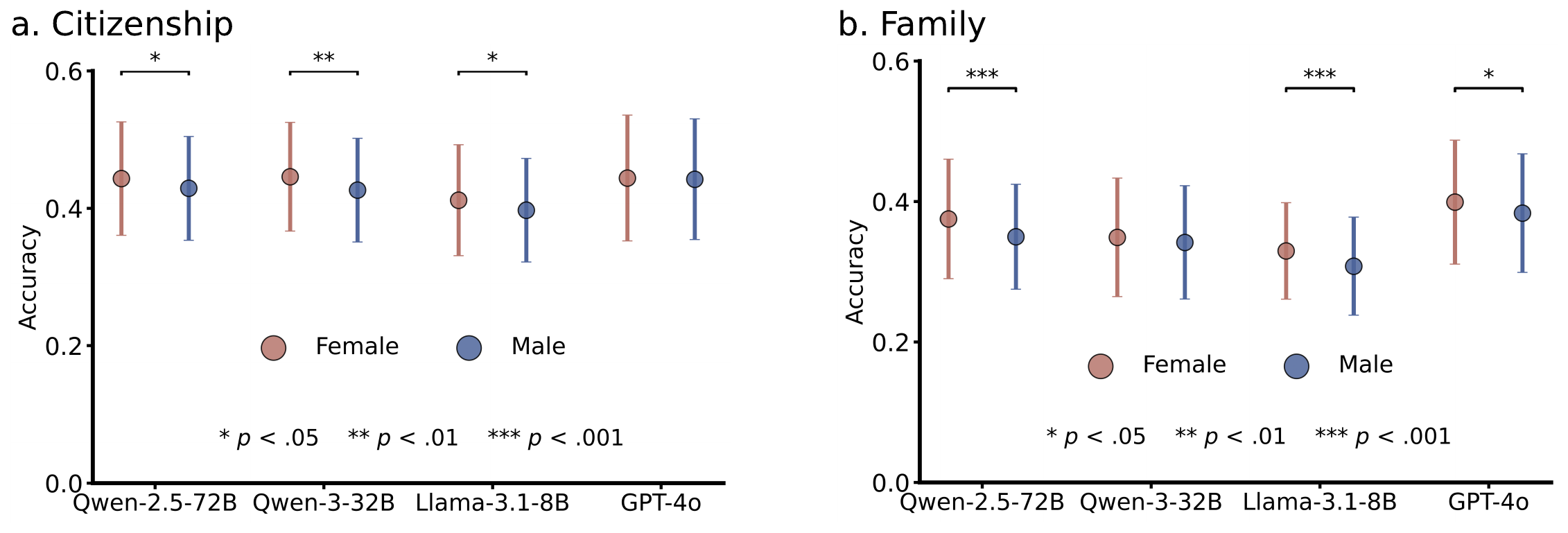}
        \caption{Experimental Results and Significance Analysis of Representative LLMs in the Cross-Gender Subgroup.}
        \label{fig:Genders_Accuracy_Comparison}
    \end{figure*}

\begin{figure*}[htbp]  
      \centering
      \includegraphics[width=.95\textwidth]{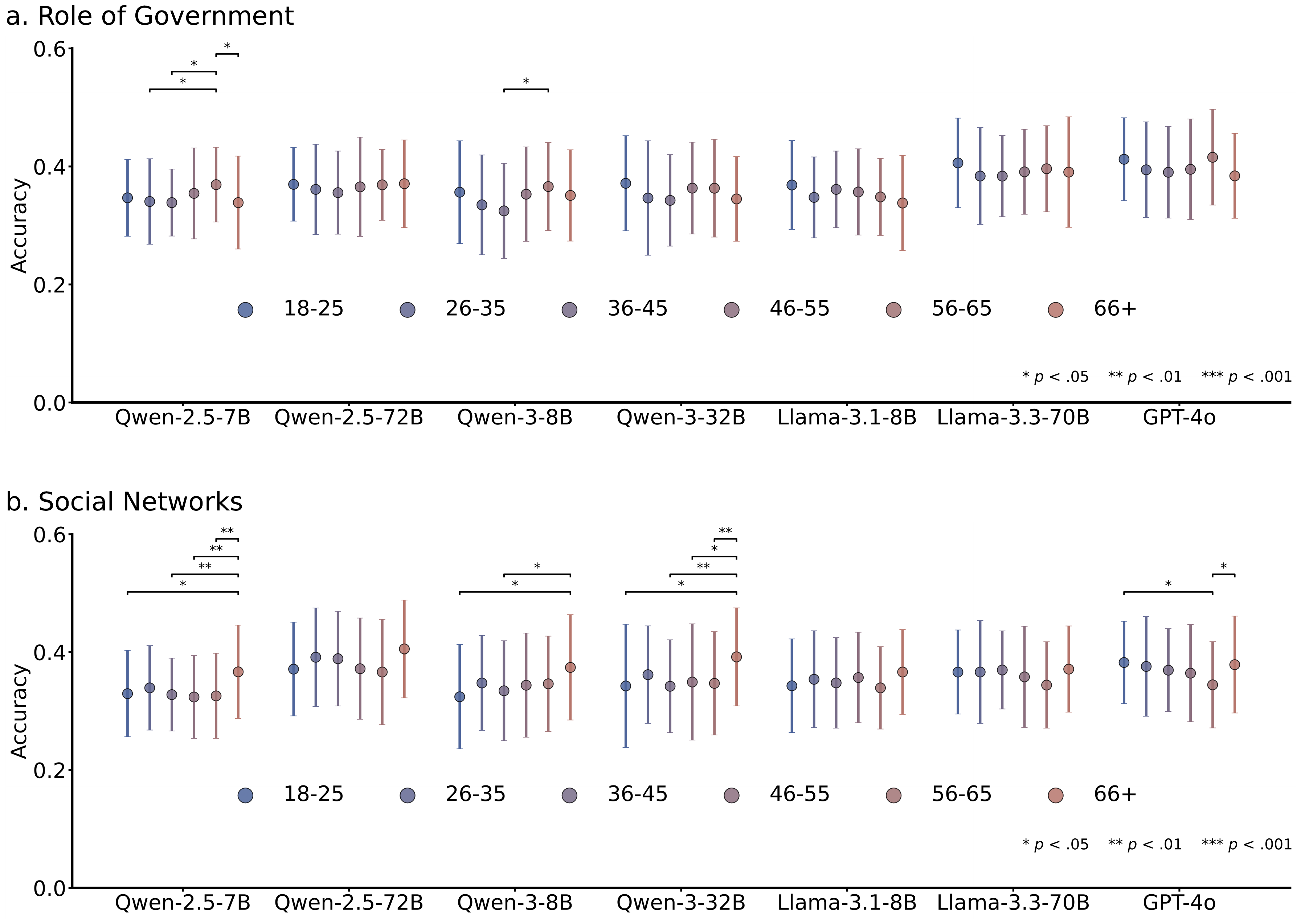}
      \caption{Experimental Results and Significance Analysis of Representative LLMs in the Cross-Age Subgroup.}
      \label{fig:Ages_Accuracy_Comparison}
    \end{figure*}

For the results of subgroup analyses by gender and age, please refer to the Figure~\ref{fig:Genders_Accuracy_Comparison} and Figure~\ref{fig:Ages_Accuracy_Comparison}.
    
\clearpage
\section{Detailed Accuracy by Demographic Variables}
\label{sec:appendix_demo}
This appendix presents the detailed model accuracy results across different demographic subgroups, broken down by domain and variable.

\begin{table}[!htbp]
\centering
\caption{Mean Accuracy (\% $\pm$ SD) of Each Model across Regions for All Domains.}
\label{tab:gender_models}
\scriptsize
\setlength{\tabcolsep}{3pt}
\begin{tabular}{llccccccc}
\toprule
\textbf{Domain} & \textbf{Continent} & Qwen2.5-7B & Qwen2.5-72B & Qwen3-8B & Qwen3-32B & Llama-3.1-8B & Llama-3.3-70B & GPT-4o \\
\midrule
\textbf{Citizenship} & Africa         & 41.8 ± 7.1 & 39.9 ± 7.8 & 41.4 ± 7.6 & 42.1 ± 8.5 & 38.1 ± 8.5 & 41.6 ± 7.8 & 40.6 ± 7.9 \\
& Asia           & 37.8 ± 10.1 & 39.8 ± 8.9 & 39.4 ± 8.4 & 41.1 ± 9.4 & 37.2 ± 9.5 & 39.4 ± 9.6 & 41.5 ± 10.3 \\
& Europe         & 41.2 ± 8.5 & 44.6 ± 7.8 & 40.4 ± 8.1 & 44.1 ± 7.5 & 41.0 ± 7.3 & 44.9 ± 8.5 & 45.0 ± 8.6 \\
& North America  & 37.6 ± 7.6 & 42.6 ± 7.4 & 37.9 ± 8.6 & 40.1 ± 5.8 & 38.4 ± 6.2 & 44.0 ± 7.9 & 46.4 ± 10.8 \\
& Oceania        & 44.0 ± 7.3 & 46.0 ± 5.6 & 42.0 ± 8.5 & 44.6 ± 4.5 & 44.3 ± 7.6 & 49.8 ± 5.9 & 48.2 ± 5.4 \\
& South America  & 45.0 ± 6.0 & 43.3 ± 5.0 & 41.4 ± 5.7 & 46.3 ± 6.1 & 42.3 ± 8.5 & 44.1 ± 7.1 & 43.2 ± 7.2 \\
\midrule
\textbf{Environment} & Africa         & 28.1 ± 6.8 & 31.1 ± 9.0 & 35.4 ± 8.1 & 30.0 ± 6.6 & 29.9 ± 6.9 & 36.4 ± 7.5 & 34.82 ± 5.8 \\
& Asia           & 30.4 ± 7.1 & 35.1 ± 7.2 & 33.4 ± 7.0 & 33.3 ± 7.3 & 33.0 ± 6.9 & 36.4 ± 7.1 & 35.9 ± 8.5 \\
& Europe         & 29.7 ± 7.6 & 36.1 ± 8.0 & 32.4 ± 7.7 & 34.5 ± 7.8 & 31.9 ± 6.8 & 35.5 ± 8.3 & 37.6 ± 8.5 \\
& North America  & 28.0 ± 7.2 & 36.7 ± 8.4 & 34.2 ± 5.6 & 35.6 ± 10.2 & 32.9 ± 8.3 & 35.6 ± 9.8 & 38.6 ± 10.3 \\
& Oceania        & 30.9 ± 6.1 & 33.8 ± 8.1 & 30.7 ± 6.2 & 34.8 ± 7.9 & 31.4 ± 7.0 & 38.0 ± 8.0 & 36.8 ± 8.9 \\
\midrule
\textbf{Family} & Africa         & 30.9 ± 4.6 & 31.2 ± 8.8 & 33.5 ± 8.0 & 29.8 ± 8.5 & 31.9 ± 7.6 & 39.6 ± 9.2 & 38.3 ± 6.7 \\
& Asia           & 28.8 ± 6.8 & 34.0 ± 8.8 & 31.6 ± 8.2 & 34.2 ± 8.4 & 30.8 ± 8.1 & 36.9 ± 9.4 & 35.6 ± 9.5 \\
& Europe         & 30.9 ± 6.8 & 37.6 ± 7.4 & 33.8 ± 7.4 & 35.5 ± 8.0 & 39.3 ± 8.3 & 39.3 ± 8.3 & 40.6 ± 8.3 \\
& North America  & 28.0 ± 5.4 & 35.2 ± 7.7 & 31.6 ± 6.9 & 32.1 ± 7.6 & 29.9 ± 5.0 & 36.8 ± 7.3 & 37.9 ± 6.6 \\
& Oceania        & 33.3 ± 4.8 & 41.5 ± 6.0 & 33.5 ± 7.4 & 35.2 ± 8.8 & 28.9 ± 7.5 & 40.9 ± 6.4 & 44.0 ± 8.5 \\
& South America  & 27.7 ± 6.5 & 32.2 ± 9.1 & 31.6 ± 7.8 & 31.2 ± 9.0 & 29.3 ± 6.2 & 37.7 ± 8.8 & 35.8 ± 8.7 \\
\midrule
\textbf{Health} & Africa         & 29.9 ± 6.4 & 34.4 ± 5.8 & 31.8 ± 7.4 & 28.1 ± 4.4 & 27.5 ± 7.0 & 30.2 ± 8.0 & 30.2 ± 6.4 \\
& Asia           & 31.7 ± 6.6 & 35.9 ± 8.7 & 32.9 ± 8.1 & 32.3 ± 7.7 & 32.4 ± 7.0 & 36.3 ± 8.7 & 35.1 ± 9.1 \\
& Europe         & 32.5 ± 7.0 & 36.6 ± 8.0 & 34.8 ± 7.4 & 34.8 ± 7.4 & 32.5 ± 7.7 & 36.7 ± 9.0 & 36.2 ± 9.4 \\
& North America  & 27.8 ± 9.3 & 30.4 ± 8.9 & 31.3 ± 7.8 & 28.3 ± 7.9 & 30.7 ± 8.2 & 32.6 ± 7.5 & 31.0 ± 8.5 \\
& Oceania        & 31.2 ± 7.8 & 36.4 ± 8.2 & 34.5 ± 8.7 & 35.2 ± 8.0 & 33.1 ± 8.7 & 37.4 ± 8.4 & 35.8 ± 7.8 \\
\midrule
\textbf{National Identity} & Asia           & 34.3 ± 7.6 & 32.3 ± 7.8 & 31.2 ± 8.5 & 29.9 ± 8.8 & 33.8 ± 7.8 & 37.9 ± 9.0 & 35.2 ± 8.8 \\
& Europe         & 33.6 ± 8.2 & 34.8 ± 8.1 & 33.7 ± 7.6 & 33.5 ± 7.5 & 33.4 ± 8.2 & 38.3 ± 8.1 & 36.7 ± 8.2 \\
& North America  & 32.4 ± 9.9 & 31.7 ± 6.3 & 32.2 ± 7.3 & 30.9 ± 7.6 & 32.0 ± 8.0 & 38.3 ± 9.4 & 35.3 ± 9.1 \\
\midrule
\textbf{Religion} & Africa         & 27.1 ± 9.7 & 31.5 ± 11.7 & 28.0 ± 10.4 & 29.3 ± 10.3 & 27.2 ± 8.2 & 31.7 ± 9.0 & 34.4 ± 10.6 \\
& Asia           & 34.4 ± 7.9 & 35.9 ± 7.9  & 36.7 ± 6.7  & 36.6 ± 8.5  & 35.1 ± 7.2 & 39.7 ± 9.0 & 38.5 ± 9.2 \\
& Europe         & 38.4 ± 8.2 & 42.0 ± 8.8  & 38.7 ± 8.7  & 40.6 ± 8.2  & 38.8 ± 8.1 & 42.7 ± 9.3 & 42.5 ± 8.6 \\
& North America  & 31.4 ± 9.1 & 37.7 ± 8.3  & 36.3 ± 8.9  & 36.7 ± 9.9  & 32.5 ± 9.5 & 37.8 ± 9.1 & 38.3 ± 11.5 \\
& South America  & 31.4 ± 7.7 & 34.6 ± 9.8  & 33.9 ± 7.9  & 34.6 ± 8.5  & 31.0 ± 7.8 & 37.9 ± 8.1 & 34.3 ± 9.9 \\
\midrule
\textbf{Role of Government} & Africa         & 33.4 ± 8.6 & 34.8 ± 4.5 & 33.6 ± 6.1 & 32.8 ± 6.7 & 32.9 ± 5.7 & 36.1 ± 4.9 & 36.5 ± 6.2 \\
& Asia           & 32.8 ± 7.3 & 34.1 ± 7.0 & 32.6 ± 8.0 & 33.1 ± 8.3 & 33.1 ± 6.8 & 36.5 ± 7.2 & 36.8 ± 7.3 \\
& Europe         & 35.3 ± 6.7 & 37.6 ± 7.3 & 35.9 ± 8.3 & 36.9 ± 8.3 & 35.9 ± 7.3 & 40.6 ± 8.1 & 41.1 ± 8.0 \\
& North America  & 36.7 ± 6.1 & 35.4 ± 4.7 & 34.2 ± 6.0 & 38.0 ± 5.5 & 35.9 ± 7.6 & 37.8 ± 8.9 & 38.8 ± 6.8 \\
& Oceania        & 38.2 ± 8.0 & 37.3 ± 8.7 & 35.5 ± 5.3 & 35.6 ± 6.9 & 33.7 ± 6.2 & 39.7 ± 7.1 & 42.4 ± 6.2 \\
& South America  & 34.4 ± 7.2 & 35.5 ± 6.8 & 31.1 ± 7.9 & 32.1 ± 7.7 & 36.9 ± 6.8 & 37.3 ± 6.4 & 39.0 ± 8.2 \\
\midrule
\textbf{Social Inequality} & Africa         & 23.6 ± 8.4 & 28.5 ± 9.1 & 26.3 ± 8.0 & 25.4 ± 6.5 & 27.4 ± 6.9 & 26.4 ± 6.8 & 29.6 ± 9.6 \\
& Asia           & 30.2 ± 9.0 & 34.2 ± 8.2 & 29.3 ± 8.8 & 31.7 ± 7.8 & 30.6 ± 8.4 & 34.0 ± 8.2 & 36.0 ± 9.2 \\
& Europe         & 31.0 ± 8.2 & 36.4 ± 9.2 & 31.8 ± 9.0 & 34.6 ± 9.0 & 32.2 ± 8.6 & 37.1 ± 9.1 & 37.8 ± 10.2 \\
& North America  & 31.3 ± 7.3 & 35.4 ± 7.9 & 31.6 ± 9.5 & 32.2 ± 8.7 & 29.7 ± 8.1 & 36.0 ± 9.2 & 37.6 ± 9.5 \\
& Oceania        & 33.2 ± 7.4 & 37.9 ± 6.7 & 34.4 ± 6.7 & 35.5 ± 8.0 & 34.1 ± 8.0 & 37.0 ± 8.9 & 39.6 ± 8.5 \\
& South America  & 27.9 ± 7.6 & 31.2 ± 8.2 & 27.4 ± 8.3 & 29.7 ± 8.5 & 29.3 ± 7.9 & 34.0 ± 9.0 & 31.8 ± 9.6 \\
\midrule
\textbf{Social Networks} & Africa         & 33.9 ± 7.2 & 39.6 ± 7.1 & 35.6 ± 7.8 & 39.3 ± 9.7 & 32.3 ± 7.9 & 35.9 ± 9.6 & 37.4 ± 8.6\\
& Asia           & 34.0 ± 7.3 & 38.3 ± 9.1 & 35.4 ± 9.7 & 36.4 ± 10.3 & 35.0 ± 8.2 & 35.7 ± 8.5 & 37.2 ± 8.8\\
& Europe         & 33.3 ± 7.3 & 37.9 ± 8.6 & 33.9 ± 8.5 & 34.7 ± 8.8 & 35.1 ± 7.5 & 36.6 ± 7.3 & 36.6 ± 7.4\\
& North America  & 32.0 ± 6.3 & 38.5 ± 8.0 & 35.2 ± 6.9 & 36.7 ± 7.3 & 35.8 ± 7.6 & 35.9 ± 7.0 & 36.3 ± 7.8\\
& Oceania        & 33.1 ± 6.8 & 38.4 ± 6.7 & 34.6 ± 8.0 & 34.6 ± 7.5 & 35.0 ± 7.0 & 36.6 ± 6.0 & 37.4 ± 6.8\\
& South America  & 31.6 ± 7.4 & 35.9 ± 8.2 & 31.8 ± 7.8 & 31.2 ± 9.0 & 33.7 ± 7.0 & 31.4 ± 6.2 & 33.4 ± 6.1\\
\midrule
\textbf{Work Orientations} & Africa         & 31.3 ± 9.0 & 36.0 ± 8.0 & 33.2 ± 6.0 & 35.7 ± 5.9 & 31.5 ± 9.0 & 38.8 ± 6.6 \\
& Asia           & 30.9 ± 7.8 & 36.3 ± 7.6 & 33.6 ± 7.6 & 34.6 ± 7.6 & 32.4 ± 6.9 & 37.0 ± 7.7 & 38.0 ± 9.1 \\
& Europe         & 32.7 ± 6.6 & 38.1 ± 7.4 & 34.7 ± 7.1 & 35.7 ± 7.6 & 33.6 ± 6.2 & 38.8 ± 7.6 & 39.5 ± 7.8 \\
& North America  & 33.7 ± 5.7 & 35.2 ± 6.5 & 33.7 ± 6.4 & 33.5 ± 7.9 & 35.4 ± 5.9 & 42.4 ± 7.9 & 36.8 ± 7.4 \\
& Oceania        & 32.6 ± 6.7 & 39.0 ± 7.0 & 34.2 ± 7.8 & 34.4 ± 7.0 & 34.3 ± 7.6 & 40.5 ± 8.2 & 42.0 ± 6.8 \\
& South America  & 29.4 ± 6.3 & 34.4 ± 7.0 & 32.1 ± 9.1 & 34.2 ± 8.1 & 32.4 ± 7.6 & 38.5 ± 7.0 & 35.1 ± 7.2 \\
\bottomrule
\end{tabular}
\end{table}

\begin{table}[htbp]
\centering
\caption{Mean accuracy (\% $\pm$ SD) of each model across gender groups for all domains.}
\label{tab:gender_models}
\scriptsize
\setlength{\tabcolsep}{4pt}  
\begin{tabular}{llccccccc}
\toprule
\textbf{Domain} & \textbf{Gender} & Qwen2.5-7B & Qwen2.5-72B & Qwen3-8B & Qwen3-32B & Llama-3.1-8B & Llama-3.3-70B & GPT-4o  \\
\midrule
\textbf{Citizenship}        & Female & 41.9 ± 9.0 & 44.3 ± 8.2 & 40.7 ± 8.3 & 44.6 ± 7.9 & 41.2 ± 8.1 & 44.2 ± 8.5 & 44.4 ± 9.2 \\
                            & Male   & 39.9 ± 8.3 & 42.9 ± 7.6 & 39.9 ± 7.8 & 42.6 ± 7.5 & 40.0 ± 7.5 & 43.8 ± 9.0 & 44.2 ± 8.8 \\
\midrule
\textbf{Environment}        & Female & 30.9 ± 7.0 & 36.2 ± 7.4 & 33.1 ± 7.7 & 34.4 ± 7.9 & 32.8 ± 7.0 & 36.6 ± 7.7 & 38.1 ± 7.5 \\
                            & Male   & 28.8 ± 7.5 & 34.8 ± 8.4 & 32.3 ± 7.0 & 33.8 ± 7.7 & 31.3 ± 6.5 & 35.3 ± 8.3 & 36.0 ± 9.3 \\
\midrule
\textbf{Family}             & Female & 30.9 ± 6.9 & 37.5 ± 8.5 & 33.7 ± 7.6 & 34.9 ± 8.4 & 33.0 ± 6.9 & 38.6 ± 8.7 & 39.9 ± 8.8 \\
                            & Male   & 29.3 ± 6.4 & 35.0 ± 7.5 & 32.5 ± 7.5 & 34.2 ± 8.1 & 30.8 ± 7.0 & 38.6 ± 8.3 & 38.3 ± 8.4 \\
\midrule
\textbf{Health}             & Female & 31.5 ± 6.8 & 35.9 ± 8.4 & 34.4 ± 7.6 & 33.7 ± 7.6 & 31.5 ± 7.5 & 36.0 ± 8.5 & 35.1 ± 9.2 \\
                            & Male   & 32.2 ± 7.6 & 35.9 ± 8.2 & 33.5 ± 7.9 & 33.5 ± 7.9 & 33.0 ± 7.8 & 36.4 ± 9.2 & 35.6 ± 9.3 \\
\midrule
\textbf{National Identity}  & Female & 33.3 ± 7.7 & 34.2 ± 8.5 & 33.2 ± 7.7 & 32.3 ± 7.9 & 33.4 ± 8.1 & 37.7 ± 8.2 & 35.8 ± 8.4 \\
                            & Male   & 34.0 ± 8.5 & 34.1 ± 7.6 & 33.0 ± 7.9 & 33.0 ± 8.0 & 33.4 ± 8.2 & 38.6 ± 8.4 & 36.8 ± 8.4 \\
\midrule
\textbf{Religion}           & Female & 36.6 ± 8.5 & 40.0 ± 9.4 & 38.2 ± 8.2 & 38.8 ± 8.2 & 37.4 ± 8.2 & 41.3 ± 9.2 & 41.0 ± 8.8 \\
                            & Male   & 36.4 ± 8.9 & 39.6 ± 9.3 & 36.9 ± 9.0 & 39.0 ± 9.4 & 36.5 ± 8.8 & 41.2 ± 9.7 & 40.5 ± 9.8 \\
\midrule
\textbf{Role of Government} & Female & 35.0 ± 7.1 & 36.8 ± 6.7 & 35.1 ± 8.2 & 35.8 ± 8.0 & 35.7 ± 7.4 & 39.3 ± 7.4 & 40.1 ± 8.0 \\
                            & Male   & 34.7 ± 7.0 & 36.3 ± 7.7 & 34.2 ± 8.1 & 35.3 ± 8.5 & 34.9 ± 7.0 & 39.1 ± 8.3 & 39.6 ± 8.0 \\
\midrule
\textbf{Social Inequality}  & Female & 30.3 ± 8.7 & 35.0 ± 9.2 & 30.6 ± 9.4 & 33.1 ± 9.0 & 31.6 ± 8.5 & 35.6 ± 8.9 & 35.9 ± 10.1 \\
                            & Male   & 30.5 ± 8.0 & 35.3 ± 8.7 & 31.1 ± 8.3 & 33.3 ± 8.8 & 31.3 ± 8.3 & 35.8 ± 9.3 & 37.3 ± 10.0 \\
\midrule
\textbf{Social Networks}    & Female & 33.5 ± 7.4 & 38.7 ± 8.7 & 34.4 ± 8.6 & 35.6 ± 8.8 & 35.0 ± 7.6 & 36.5 ± 7.1 & 37.1 ± 7.8 \\
                            & Male   & 33.2 ± 7.0 & 37.4 ± 8.2 & 34.4 ± 8.7 & 34.9 ± 9.4 & 35.0 ± 7.7 & 35.8 ± 8.1 & 36.2 ± 7.7 \\
\midrule
\textbf{Work Orientations}  & Female & 32.4 ± 6.9 & 37.4 ± 7.3 & 34.2 ± 7.2 & 35.5 ± 8.0 & 33.7 ± 6.5 & 38.8 ± 7.6 & 38.7 ± 7.9 \\
                            & Male   & 31.9 ± 7.0 & 37.4 ± 7.5 & 34.2 ± 7.4 & 35.0 ± 7.1 & 33.1 ± 6.6 & 38.8 ± 7.6 & 39.1 ± 8.1 \\
\bottomrule
\end{tabular}
\end{table}

\begin{table}[htbp]
\centering
\caption{Mean accuracy (\% $\pm$ SD) of each model across age ranges for all domains.}
\label{tab:age_models}
\scriptsize
\setlength{\tabcolsep}{3.5pt}  
\begin{tabular}{llccccccc}
\toprule
\textbf{Domain} & \textbf{Age Range} & Qwen2.5-7B & Qwen2.5-72B & Qwen3-8B & Qwen3-32B & Llama-3.1-8B & Llama-3.3-70B & GPT-4o \\
\midrule
\textbf{Citizenship} & 18–25 & 41.5 ± 8.4 & 45.2 ± 7.1 & 40.5 ± 7.9 & 43.6 ± 8.6 & 39.8 ± 8.2 & 45.1 ± 8.6 & 44.0 ± 9.4 \\
                     & 26–35 & 40.2 ± 9.7 & 42.8 ± 7.4 & 39.9 ± 8.2 & 44.4 ± 8.0 & 39.8 ± 7.7 & 42.6 ± 8.4 & 43.3 ± 8.9 \\
                     & 36–45 & 40.7 ± 10.1 & 43.4 ± 8.9 & 40.2 ± 8.5 & 43.9 ± 7.7 & 39.8 ± 8.4 & 44.4 ± 8.7 & 43.7 ± 8.8 \\
                     & 46–55 & 39.3 ± 7.5 & 42.8 ± 7.6 & 40.7 ± 8.2 & 43.0 ± 7.2 & 40.5 ± 7.3 & 43.8 ± 8.5 & 43.6 ± 8.1 \\
                     & 56–65 & 42.0 ± 8.4 & 43.6 ± 7.8 & 39.6 ± 6.8 & 42.5 ± 8.4 & 40.9 ± 7.7 & 43.7 ± 9.4 & 45.2 ± 9.7 \\
                     & 66+   & 42.5 ± 7.9 & 44.7 ± 8.2 & 40.9 ± 8.3 & 44.5 ± 7.5 & 41.4 ± 8.0 & 44.8 ± 9.0 & 46.0 ± 9.3 \\
\midrule
\textbf{Environment} & 18–25 & 32.1 ± 6.4 & 33.9 ± 6.8 & 31.8 ± 5.9 & 32.7 ± 7.1 & 33.0 ± 6.8 & 36.7 ± 7.0 & 38.0 ± 8.2 \\
                     & 26–35 & 30.2 ± 7.3 & 35.1 ± 8.6 & 32.2 ± 8.0 & 33.5 ± 8.8 & 31.9 ± 6.9 & 34.8 ± 8.4 & 36.9 ± 10.5 \\
                     & 36–45 & 28.2 ± 7.1 & 35.1 ± 7.3 & 33.9 ± 7.2 & 34.8 ± 6.9 & 31.5 ± 7.6 & 35.1 ± 7.5 & 36.9 ± 7.7 \\
                     & 46–55 & 30.0 ± 6.8 & 35.5 ± 7.6 & 32.4 ± 7.4 & 34.0 ± 8.1 & 32.2 ± 6.3 & 36.3 ± 8.0 & 36.4 ± 8.6 \\
                     & 56–65 & 30.0 ± 7.3 & 36.3 ± 8.0 & 32.8 ± 7.5 & 33.6 ± 7.4 & 32.3 ± 6.9 & 36.1 ± 8.8 & 37.3 ± 7.8 \\
                     & 66+   & 29.7 ± 8.2 & 36.1 ± 8.1 & 32.8 ± 7.5 & 35.7 ± 8.1 & 31.9 ± 6.7 & 36.9 ± 7.2 & 37.6 ± 8.0 \\
\midrule
\textbf{Family}      & 18–25 & 29.6 ± 3.9 & 35.1 ± 7.4 & 30.7 ± 8.7 & 31.3 ± 11.1 & 31.9 ± 6.6 & 39.6 ± 9.2 & 38.6 ± 10.1 \\
                     & 26–35 & 29.9 ± 6.9 & 34.3 ± 8.6 & 31.4 ± 6.4 & 32.2 ± 8.5 & 31.2 ± 7.1 & 35.8 ± 7.3 & 37.0 ± 8.3 \\
                     & 36–45 & 31.4 ± 6.5 & 36.3 ± 8.4 & 33.5 ± 8.2 & 34.4 ± 8.9 & 32.1 ± 7.2 & 38.3 ± 9.2 & 38.5 ± 9.0 \\
                     & 46–55 & 29.6 ± 7.0 & 35.6 ± 7.7 & 33.7 ± 8.1 & 33.9 ± 8.2 & 31.9 ± 7.0 & 38.6 ± 8.1 & 38.8 ± 8.8 \\
                     & 56–65 & 29.1 ± 6.8 & 36.7 ± 7.6 & 32.6 ± 7.2 & 35.5 ± 7.7 & 31.5 ± 6.5 & 38.2 ± 7.8 & 40.2 ± 8.6 \\
                     & 66+   & 30.6 ± 6.4 & 38.2 ± 8.5 & 33.5 ± 7.0 & 36.2 ± 7.2 & 32.6 ± 7.6 & 41.6 ± 9.0 & 40.6 ± 8.0 \\
\midrule
\textbf{Health}      & 18–25 & 31.9 ± 8.1 & 36.2 ± 7.8 & 32.9 ± 7.2 & 34.0 ± 8.1 & 31.3 ± 7.0 & 38.5 ± 8.2 & 34.9 ± 9.2 \\
                     & 26–35 & 30.9 ± 7.3 & 35.9 ± 7.7 & 33.0 ± 8.4 & 33.3 ± 7.9 & 31.8 ± 7.4 & 37.4 ± 8.7 & 34.3 ± 9.5 \\
                     & 36–45 & 30.8 ± 7.2 & 35.2 ± 8.1 & 34.0 ± 7.4 & 33.8 ± 7.7 & 31.4 ± 7.7 & 36.4 ± 8.8 & 36.0 ± 9.5 \\
                     & 46–55 & 27.8 ± 9.3 & 32.5 ± 8.8 & 32.0 ± 8.0 & 31.1 ± 7.4 & 29.5 ± 8.2 & 34.3 ± 9.0 & 34.9 ± 9.4 \\
                     & 56–65 & 31.2 ± 7.8 & 36.4 ± 8.2 & 34.5 ± 8.7 & 35.2 ± 8.0 & 32.5 ± 7.9 & 37.5 ± 9.3 & 34.5 ± 8.7 \\
                     & 66+   & 32.9 ± 8.4 & 37.3 ± 8.5 & 35.4 ± 9.0 & 35.5 ± 8.7 & 34.6 ± 8.5 & 38.6 ± 9.5 & 37.1 ± 8.9 \\
\midrule
\textbf{National Identity} & 18–25 & 33.3 ± 7.7 & 34.7 ± 8.1 & 32.4 ± 7.9 & 31.7 ± 8.2 & 33.7 ± 8.0 & 38.5 ± 7.8 & 33.3 ± 8.3 \\
                           & 26–35 & 34.2 ± 7.8 & 34.9 ± 8.0 & 32.7 ± 8.3 & 32.5 ± 8.0 & 33.7 ± 8.1 & 39.2 ± 8.3 & 35.9 ± 8.5 \\
                           & 36–45 & 33.7 ± 8.3 & 35.5 ± 7.9 & 33.3 ± 7.5 & 32.8 ± 7.4 & 33.9 ± 7.6 & 39.4 ± 8.4 & 38.0 ± 7.3 \\
                           & 46–55 & 32.4 ± 9.9 & 34.8 ± 8.4 & 32.4 ± 7.7 & 32.1 ± 7.8 & 33.3 ± 7.8 & 38.8 ± 9.2 & 36.9 ± 9.0 \\
                           & 56–65 & 34.0 ± 7.8 & 35.6 ± 8.2 & 33.6 ± 7.9 & 33.1 ± 8.0 & 33.8 ± 7.9 & 38.2 ± 8.5 & 36.0 ± 8.3 \\
                           & 66+   & 34.4 ± 8.4 & 35.9 ± 8.0 & 33.3 ± 8.6 & 32.8 ± 8.6 & 34.5 ± 8.1 & 39.5 ± 9.0 & 37.2 ± 8.2 \\
\midrule
\textbf{Religion} & 18--25 & 34.7 ± 9.7 & 38.9 ± 10.7 & 35.4 ± 8.4 & 37.5 ± 10.1 & 35.5 ± 8.5 & 39.7 ± 9.9 & 39.6 ± 10.4 \\
& 26--35 & 35.8 ± 7.9 & 38.0 ± 9.2  & 37.3 ± 9.4 & 37.9 ± 8.9  & 35.2 ± 7.9 & 41.8 ± 9.3 & 39.4 ± 9.2 \\
& 36--45 & 36.9 ± 8.0 & 40.3 ± 8.7  & 37.5 ± 8.2 & 39.2 ± 8.3  & 38.1 ± 7.9 & 41.5 ± 8.8 & 41.7 ± 9.5 \\
& 46--55 & 36.6 ± 8.2 & 40.2 ± 8.8  & 37.7 ± 8.6 & 38.1 ± 8.4  & 37.7 ± 9.3 & 41.3 ± 9.3 & 41.4 ± 8.0 \\
& 56--65 & 38.2 ± 9.9 & 40.9 ± 10.1 & 38.8 ± 8.9 & 40.5 ± 9.2  & 38.2 ± 8.7 & 42.6 ± 9.4 & 42.1 ± 10.2 \\
& 66+    & 35.9 ± 8.3 & 39.6 ± 8.7  & 38.0 ± 8.0 & 39.6 ± 8.1  & 36.0 ± 8.0 & 39.8 ± 10.2 & 39.1 ± 8.3 \\
\midrule
\textbf{Role of Government} & 18--25 & 34.7 ± 6.5 & 37.0 ± 6.2 & 35.6 ± 8.7 & 37.1 ± 8.1 & 36.8 ± 7.5 & 40.6 ± 7.6 & 41.2 ± 7.0 \\
& 26--35 & 34.1 ± 7.2 & 36.1 ± 7.6 & 33.5 ± 8.5 & 34.6 ± 9.7 & 34.8 ± 6.9 & 38.4 ± 8.2 & 39.4 ± 8.1 \\
& 36--45 & 33.9 ± 5.7 & 35.6 ± 7.0 & 32.5 ± 8.1 & 34.2 ± 7.8 & 36.1 ± 6.5 & 38.4 ± 6.9 & 39.0 ± 7.8 \\
& 46--55 & 35.4 ± 7.7 & 36.5 ± 8.4 & 35.3 ± 8.0 & 36.3 ± 7.8 & 35.7 ± 7.3 & 39.1 ± 7.2 & 39.5 ± 8.5 \\
& 56--65 & 36.9 ± 6.3 & 36.9 ± 6.0 & 36.6 ± 7.5 & 36.3 ± 8.3 & 34.8 ± 6.5 & 39.6 ± 7.3 & 41.6 ± 8.1 \\
& 66+    & 33.9 ± 7.9 & 37.1 ± 7.4 & 35.1 ± 7.7 & 34.5 ± 7.2 & 33.8 ± 8.0 & 39.0 ± 9.4 & 38.4 ± 7.2 \\
\midrule
\textbf{Social Inequality} & 18--25 & 28.5 ± 8.3 & 34.8 ± 7.6 & 30.6 ± 6.8 & 33.6 ± 7.8 & 30.3 ± 7.8 & 34.9 ± 6.7 & 34.9 ± 11.1 \\
& 26--35 & 30.6 ± 7.6 & 35.2 ± 9.3 & 30.8 ± 7.6 & 33.7 ± 7.5 & 31.8 ± 7.8 & 33.6 ± 8.3 & 35.5 ± 8.0 \\
& 36--45 & 30.6 ± 8.4 & 35.2 ± 9.0 & 31.2 ± 8.7 & 33.2 ± 8.6 & 31.3 ± 8.9 & 36.8 ± 9.3 & 38.0 ± 10.4 \\
& 46--55 & 30.2 ± 8.8 & 35.3 ± 9.0 & 31.3 ± 8.5 & 33.1 ± 8.3 & 31.7 ± 7.8 & 36.7 ± 9.6 & 37.1 ± 10.5 \\
& 56--65 & 31.5 ± 8.7 & 35.1 ± 9.3 & 30.9 ± 9.8 & 33.2 ± 9.6 & 31.7 ± 9.2 & 34.9 ± 9.6 & 36.1 ± 10.6 \\
& 66+    & 30.0 ± 8.0 & 35.2 ± 9.0 & 30.5 ± 10.2 & 32.8 ± 10.6 & 31.4 ± 8.8 & 36.6 ± 9.2 & 36.9 ± 9.8 \\
\midrule
\textbf{Social Networks} & 18--25 & 33.0 ± 7.3 & 37.1 ± 8.0 & 32.4 ± 8.8 & 34.3 ± 10.4 & 34.3 ± 7.9 & 36.6 ± 7.1 & 38.2 ± 7.0 \\
& 26--35 & 33.9 ± 7.2 & 39.1 ± 8.3 & 34.8 ± 8.0 & 36.2 ± 8.3  & 35.4 ± 8.2 & 36.6 ± 8.7 & 37.6 ± 8.5 \\
& 36--45 & 32.8 ± 6.2 & 38.9 ± 8.0 & 33.4 ± 8.5 & 34.2 ± 7.9  & 34.8 ± 7.7 & 37.0 ± 6.6 & 36.9 ± 7.0 \\
& 46--55 & 32.4 ± 7.0 & 37.2 ± 8.6 & 34.4 ± 8.8 & 34.9 ± 9.9  & 35.7 ± 7.7 & 35.8 ± 8.6 & 36.4 ± 8.2 \\
& 56--65 & 32.6 ± 7.2 & 36.6 ± 8.9 & 34.6 ± 8.1 & 34.7 ± 8.8  & 33.9 ± 7.0 & 34.4 ± 7.3 & 34.4 ± 7.3 \\
& 66+    & 36.6 ± 7.9 & 40.5 ± 8.3 & 37.4 ± 8.9 & 39.2 ± 8.3  & 36.6 ± 7.2 & 37.1 ± 7.3 & 37.9 ± 8.2 \\
\midrule
\textbf{Work Orientations} & 18--25 & 32.9 ± 7.0 & 35.9 ± 6.5 & 33.5 ± 6.5 & 33.9 ± 8.1  & 32.4 ± 7.4 & 38.0 ± 7.0 & 37.5 ± 6.9 \\
& 26--35 & 31.2 ± 6.6 & 35.9 ± 6.8 & 33.7 ± 7.6 & 35.0 ± 8.3  & 33.1 ± 6.6 & 38.0 ± 7.4 & 38.7 ± 8.2 \\
& 36--45 & 32.1 ± 7.2 & 37.9 ± 7.5 & 33.8 ± 7.1 & 35.2 ± 7.3  & 33.4 ± 6.4 & 38.6 ± 7.7 & 39.6 ± 8.5 \\
& 46--55 & 33.2 ± 6.9 & 38.4 ± 8.0 & 35.0 ± 8.0 & 36.0 ± 7.7  & 34.6 ± 6.3 & 40.3 ± 7.9 & 39.4 ± 8.0 \\
& 56--65 & 31.8 ± 6.8 & 37.2 ± 7.2 & 34.0 ± 6.4 & 34.9 ± 7.0  & 32.3 ± 7.0 & 38.1 ± 7.2 & 38.0 ± 7.3 \\
& 66+    & 32.7 ± 6.8 & 40.1 ± 6.3 & 39.5 ± 6.6 & 37.5 ± 4.9  & 32.5 ± 4.3 & 39.7 ± 9.4 & 41.4 ± 5.0 \\
\bottomrule
\end{tabular}
\end{table}

\end{document}